\documentclass[]{aastex7}  

\usepackage{graphicx}
\usepackage{amsmath}
\usepackage{amsthm}
\usepackage{natbib}
\usepackage{diagbox}

\begin{document}

   \title{Prospects of detecting rotational flatness of exoplanets from space-based photometry}


\author[orcid=0000-0003-3754-7889]{Szil\'ard K\'alm\'an}
\affiliation{Konkoly Observatory, HUN-REN Research Centre for Astronomy and Earth Sciences, Konkoly Thege 15-17, 1121 Budapest, Hungary}
\affiliation{HUN-REN CSFK, MTA Centre of Excellence, Budapest, Hungary}
\affiliation{ELTE E{\"o}tv{\"o}s Lor\'and University, Doctoral School of Physics,  Budapest, Pázmány Péter sétány 1/A, H-1117, Hungary}
\email{kalman.szilard@csfk.org}
\author[orcid=0000-0001-6803-9698]{Szil\'ard Csizmadia}
\affiliation{German Aerospace Center (DLR), Institute of  Planetary Research, Department of Extrasolar Planets and Atmospheres, 12489 Berlin, Rutherfordstrasse 2., Berlin, Germany}
\email{szilard.csizmadia@dlr.de}
\author{Lia M. Bernab\'o}
\affiliation{German Aerospace Center (DLR), Institute of  Planetary Research, Department of Extrasolar Planets and Atmospheres, 12489 Berlin, Rutherfordstrasse 2., Berlin, Germany}
\email{lia.bernabo@dlr.de}
\author[orcid=0000-0002-3258-1909]{R\'obert Szab\'o}
\affiliation{Konkoly Observatory, HUN-REN Research Centre for Astronomy and Earth Sciences, Konkoly Thege 15-17, 1121 Budapest, Hungary}
\affiliation{HUN-REN CSFK, MTA Centre of Excellence, Budapest, Hungary}
\affiliation{ELTE E\"otv\"os Lor\'and University, Institute of Physics and Astronomy, P\'azm\'any P\'eter s\'et\'any 1/A, H-1117 Budapest, Hungary}
\email{szabo.robert@csfk.org}

\author[orcid=0000-0002-0606-7930]{Gyula Szab\'o M.}
\affiliation{ELTE E{\"o}tv{\"o}s Lor\'and University, Gothard Astrophysical Observatory, Szombathely, Szent Imre h. u. 112., H-9700, Hungary}
\affiliation{    MTA-ELTE  Lend{\"u}let "Momentum" Milky Way Research Group, Szombathely, Szent Imre h. u. 112., H-9700, Hungary}
\email{szgy@gothard.hu}

  \begin{abstract}
    In the era of photometry with space-based telescopes, such as CHEOPS (CHaracterizing ExOPlanets Satellite), JWST (James Webb Space Telescope), PLATO (PLAnetary Transits and Oscillations of stars), and ARIEL (Atmospheric Remote-sensing Infrared Exoplanet Large-survey), the road has opened for detecting subtle distortions in exoplanet transit light curves—resulting from their non-spherical shape.
   We investigate the prospects of retrieval of rotational flatness (oblateness) of exoplanets at various noise levels.
   We present a novel method for calculating the transit light curves based on the Gauss-Legendre quadrature. We compare it in the non-rotating limit to the available analytical models. We conduct injection-and-retrieval tests to assess the precision and accuracy of the retrievable oblateness values.
   We find that the light curve calculation technique is about $25$\% faster than a well-known analytical counterpart, while still being precise enough. We show that a $3 \sigma$ oblateness detection is possible for a planet orbiting bright enough stars, by exploiting a precise estimate on the stellar density obtained e.g. from asteroseismology. We also show that for noise levels $\geq 256$~ppm (expressed as point-to-point scatter with a $60$~s exposure time) detection of planetary oblateness is not reliable.
  \end{abstract}


%
\newcommand{\bigrpol}{R_{\rm Pol}}
\newcommand{\bigreq}{R_{\rm Eq}}
\newcommand{\rpol}{r_{\rm Pol}}
\newcommand{\req}{r_{\rm Eq}}
\newcommand{\massp}{M_{\rm P}}
\newcommand{\rp}{R_{\rm P}}
\section{Introduction}

Of the nearly $6000$ confirmed exoplanets known to date, more than $4000$ have been discovered by observing and analyzing their transits. This process has been primarily facilitated by the unrivaled precision of space-based telescopes, such as \textit{Kepler} \citep{2010Sci...327..977B}, \textit{TESS} \citep[Transiting Exoplanet Survey Satellite;][]{2015JATIS...1a4003R}, and \textit{CoRoT} \citep[Convection, Rotation and planetary Transits;][]{2009A&A...506..411A, deleuil18}. In the future, the \textit{PLATO} \citep[PLAnetary Transits and Oscillations of stars;][]{2014ExA....38..249R, 2024arXiv240605447R} mission is expected to continue detection of new exoplanets \citep{2023A&A...677A.133M}. Following the succesful launch of \textit{CHEOPS} \citep[CHaracterizing ExOPlanets Satellite;][]{2021ExA....51..109B}, and \textit{JWST} \citep[James Webb Space Telescope;][]{2006SSRv..123..485G, 2023PASP..135e8001M}, characterizing individual planetary systems also became feasible at an unprecedented level of precision -- a trend expected to be continued with \textit{ARIEL} \citep[Atmospheric Remote-sensing Infrared Exoplanet Large-survey;][]{2018ExA....46..135T, 2022EPSC...16.1114T} in the upcoming decade. 

The ultra-precise space-based observations allow for the observation of the non-spherical nature of certain exoplanets. These distortions in the idealized spherical shape of objects may arise as a result of either tidal interactions with the star (for close-in planets) or due to the rapid rotation of the planet (also for hot and cool planets). Detecting the tidal distortions of the planetary shape can be done indirectly by observing the tidal decay of the orbit either in radial velocities or transit timing variations \citep[e.g.][]{2019A&A...623A..45C, 2025A&A...694A.233B}, or directly by modeling the deformed transit shape \citep{2019ApJ...878..119H, 2020ApJ...889...66H, akinsamni19, 2024JOSS....9.6972C} and the phase curve \citep{2024A&A...685A..63A,2024A&A...682A..15A}\footnote{Note that the mass ratio of the primary and the secondary objects in a binary system from the ellipsoidal effect counterpart of the phase curve was shown to be determinable by \citet{russell12,wd71}}. The detection of rotational flattening (oblateness) is made possible by calculating the transit of a spheroid (i.e. the overlap between the circular stellar and elliptical planetary disk). For spherical planets, the analytical model of \cite{2002ApJ...580L.171M} is commonly used, due to its speed and precision. The case of ellipsoidal planets is more computationally demanding. \cite{2022ExA....53..607S} used a brute-force approach, where the stellar disk is calculated on an $8000 \times 8000$ pixel disk, and the flux loss during the transit is given by shifting the planetary disk as a mask across the transit chord. \cite{2024AJ....168..243B} present a general-purpose Monte Carlo tool, useful for calculating the transits of arbitrarily shaped objects. \cite{2019MNRAS.490.1111R} developed a model that approximates the boundaries with polygons and calculated the transits accordingly. \cite{2024JOSS....9.6972C, 2024arXiv240611644L,2024arXiv241003449D}, and \cite{2025ApJ...981L...7P} all constructed models that transform the two-dimensional surface integrals (where the disk of the planet and the star overlap) to line integrals using Green's theorem, which can be solved efficiently. In this paper, we present a novel method for calculating the transit light curves, based on the Gauss-Legendre quadrature \citep[see e.g.][]{trefethen2006a}. 


This paper is structured as follows. We provide the parameterization of the elliptical planetary shape as well as the basis of the integration used for the transit calculations in Sect. \ref{sec:transit}. In Sect. \ref{sec:tests}, we assess the limitations and benefits of our model by comparing it in the non-oblate limit to the analytical approach of \cite{2002ApJ...580L.171M}. We conduct a large-scale injection-and-recovery test on $800$ transit light curves with different oblateness parameters and noise properties, and present the results in Sect. \ref{sec:results}. Signal-to-noise criteria for detecting the oblateness are shown in Sect. \ref{sec:param_recovery}. We draw our conclusions in Sect. \ref{sec:conclusion}.
\section{Transits of an ellipsoidal planet} \label{sec:transit}
\subsection{Ellipsoidal planetary shape}

Let us assume that due to the rotation of the planet, its shape can be described perfectly with a biaxial ellipsoid (also known as a spheroid). In Cartesian coordinates, this can be described as
\begin{equation}
    \frac{x^2}{\mathcal{A}^2} + \frac{y^2}{\mathcal{B}^2} +\frac{z^2}{\mathcal{C}^2} = 1,
\end{equation}
where $\mathcal{A} = \mathcal{B} \geq \mathcal{C}$ and $z$ is the spin axis. During transit, the light originating from the stellar surface is obscured by the sky-projected shape of the transiting body along the transit chord. We define the polar ($\bigrpol$) and equatorial ($\bigreq$) radii of the planet based on its projected shape -- an ellipse, where $\bigrpol \leq \bigreq$. We define the oblateness of the planet via
\begin{equation}
    f = \frac{\bigreq-\bigrpol}{\bigreq}.
\end{equation}
It can be shown that $\bigreq = A$  and $\bigrpol \geq C$. 
Consequently, the observable oblateness $f$ is only a lower limit of the true oblateness, $F = \frac{A-C}{A}$ \citep[see][for the relationship between $F$ and $f$]{2009ApJ...705..683B}. The two are equal only when the inclination of the planetary spin axis with respect to the line of sight is $90^\circ$. Given that $F$ cannot be determined from photometry, we refer to $f$ as the oblateness of the planet. We define the relative equatorial and polar radii as $\rpol = \bigrpol/R_\star$ and $\req = \bigreq/R_\star$, respectively. Note that $f = \frac{\req-\rpol}{\req}$.

Neglecting tidal forces, the surface of the planet is conformed to the gravitational potential $V$ as
\begin{equation} \label{eq:potential}
    V = -\frac{GM_{\rm P}}{\rpol} = -\frac{GM_{\rm P}}{\req} - \frac{1}{2}\omega_{\rm rot}^2 \req^2,
\end{equation}
where $G$ is the gravitational constant, $M_{\rm P}$ is the planetary mass and $\omega$ is the angular velocity of the planet, which is assumed to be a rigid body. Rearranging Eq (\ref{eq:potential}), we get
\begin{equation} \label{eq:reqrpol}
     \frac{\req}{\rpol} = 1 + \frac{1}{2}\frac{\omega^2 \req^3}{GM_{\rm P}}.
\end{equation}
The break-up of a solid body occurs when the equatorial centrifugal force is greater than the gravitational force (at the equator). At the break-up velocity, the planet is rotating with the  $\omega_{\rm rot, crit}$, therefore, at the limit, we find \citep[in agreement with e.g.][]{2022ApJ...935..178B}
\begin{equation}
    \omega_{\rm rot, crit}^2 \req = \frac{G\massp}{\req^2}.
\end{equation}
Substituting the critical angular velocity, $\omega_{\rm rot, crit} = \sqrt{\frac{G\massp}{\req^3}}$, into Eq (\ref{eq:reqrpol}), we find
\begin{equation}
   \frac{\req}{\rpol}= 1+\frac{1}{2}\frac{G\massp\req^3}{G\massp\req^3} = \frac{3}{2}.
\end{equation}
Consequently, the largest possible oblateness is $f_{\rm crit} = \frac{1}{3}$. In this approach, we neglect the quadrupole moment of the planet \citep[in contrast with e.g.][]{2022ApJ...935..178B}, thus arriving at a different critical value of $f$. We also note that \cite{2022ApJ...935..178B} apparently provide a lower limit of $0.5$ on $f$ (Eq. (5) of \citealt{2022ApJ...935..178B}).

\subsection{Transit calculation} \label{sect:transit_calc}

\begin{figure}
    \centering
    \includegraphics[width=\columnwidth]{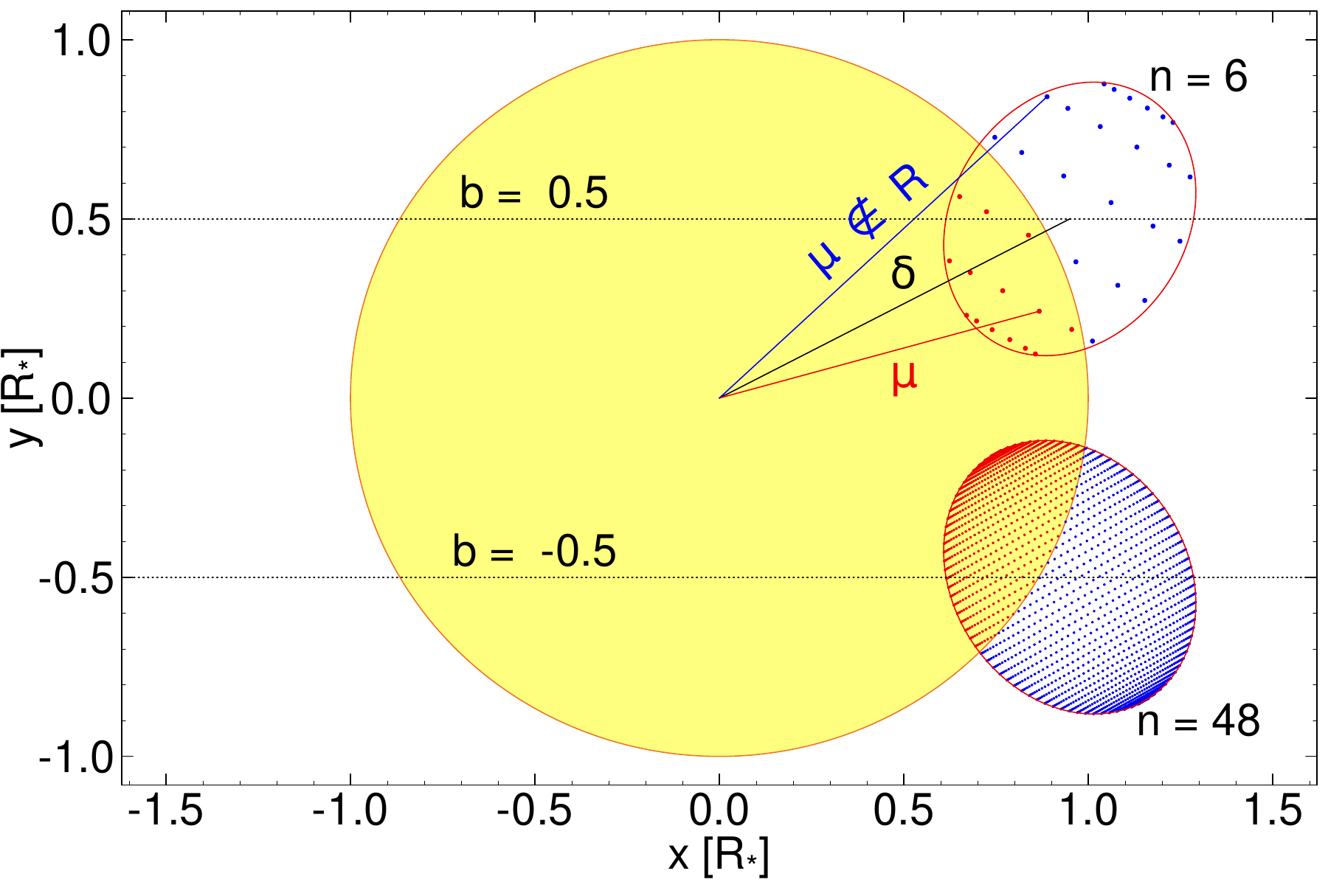}
    \caption{Gauss-Legendre quadrature points used for the transit light curve calculations of an oblate planet (red ellipsoidal contours) in front of a star (yellow disk). When $\mu \in \mathbb{R}$, the integration point overlaps with the star, thus it is included in the transit computations (red points). On the other hand, when $\mu \notin \mathbb{R}$, the (blue) point does not block light from the stellar disk. More quadrature points yield more precise light curves as contours of the overlapping areas are sampled better.}
    \label{fig:gl_transit}
\end{figure}

 The Gauss-Legendre quadrature points are defined via the zero points of the Legendre polynomials. In a one dimensional case, the finite integral of a function $\mathcal{F}(x)$ on the closed interval $[a, b]$  can be calculated as
\begin{equation} \label{eq:glq}
    \int_a^b \mathcal{F}(x) dx \approx \frac{b-a}{2} \sum_{i=1}^n w_i \mathcal{F}\left(\frac{a+b}{2} + \frac{b-a}{2}x_i \right),
\end{equation}
where $x_i$ is the $i$th root of the Legendre polynomial $P_n (x)$. The weight factors ($w_i$) are defined as
\begin{equation}
    w_i = \frac{2}{\left(1 - x_i^2 \right) \left.\left( \frac{dP_n(x)}{dx} \right)^2 \right\vert_{x = x_i}}.
\end{equation}
As an example, for $n=6$, the ($x_i$, $w_i$) pairs are listed in Table \ref{tab:gl_qp}. Note that $\sum_i  w_i = 2$, independently of the number of integration points. The integration is carried out utilizing the \verb|INT_2D| function built into IDL, which can perform the computations using $n \in \left\{6, 10, 20, 48, 96 \right\}$. 

\begin{table} \label{sec:glq}
\caption{Gauss-Legendre quadrature points for $n = 6$ points.}
    \label{tab:gl_qp}
    \centering
    \begin{tabular}{c c c}
    \hline
        $i$ & $x_i$ & $w_i$  \\
        \hline
        \hline
        1 &$- 0.932469514203152$ & $0.171324492379170$ \\
        2 &$- 0.661209386466265$ &$0.360761573048139$ \\
        3 &$- 0.238619186083197$ &$0.467913934572691$ \\
        4 &$+ 0.238619186083197$ &$0.467913934572691$ \\
        5 &$+ 0.661209386466265$ &$0.360761573048139$ \\
        6 &$+ 0.932469514203152$ &$0.171324492379170$ \\
\hline
\end{tabular}
\end{table}

In the case of the ellipsoidal planet, the shape of the transiting object is enclosed via an ellipse. The integration is carried out in two dimensions, on an $n \times n$ basis. The integral of $\mathcal{F}(x, y)$ over the closed intervals $[a, b]$ and $[c, d]$ can be written as

\begin{equation}
\begin{split} \label{eq:int2d}
    \int_{a}^b \int_c^d\mathcal{F}(x, y) dx dy  \approx \frac{(b-a)(d-c)}{4} \sum_{j=1}^n \sum_{i = 1}^n w_j w_i \mathcal{F} \left(\frac{a+b}{2} + \frac{b-a}{2}x_i, \frac{c+d}{2} + \frac{d-c}{2}y_i  \right).
\end{split}
\end{equation}

We define the coordinate system so that the $x$ axis coincides with the major axis ellipse that defines the contour of the sky-projected spheroid, while the $y$ axis coincides with the minor axis (in the orientation of the north pole of the planet). The origin of the coordinate system is the center of the boundary ellipse.

The coordinates of the integration points along the major axis of the ellipse can be represented as an $n \times n$ matrix: \begin{equation}
    \mathbf{x}_{ij} =\req \cdot
    \begin{bmatrix}
    x_1 & x_2 & \dots & x_n \\
    x_1 & x_2 & \dots & x_n \\
    \vdots & \vdots &  & \vdots \\
    x_1 & x_2 & \dots & x_n\\
    \end{bmatrix} = \req \cdot \mathbf{I} \cdot \vec{x}_i,
\end{equation}
where $\mathbf{I}$ is the unit matrix and $\vec{x}_i$ is a vector containing all $n$ roots of the Legendre polynomial $P_n (x)$\footnote{For clarity, $x_{kl}$ and $x_m$ refer to the individual elements of the matrix or the vector.} To derive the $y_{ij}$ coordinates of the integration points, we first need to find limits corresponding to every $x_i$ in the direction of the minor axis ($y_{{\rm limit},i}$). Following the equation of the ellipse
\begin{equation}
    \frac{x_{i}^2}{\req^2} + \frac{y_{{\rm limit}, i}^2}{\rpol^2} = 1,
\end{equation}
which leads to
\begin{equation}
y_{{\rm limit},i} = \req (1-f) \sqrt{1-x_i^2}.
\end{equation}
The coordinates are then given as
\begin{equation}
    \mathbf{y}_{ij} = \vec{y}_{{\rm limit}, i} \cdot \vec{y}_j.
\end{equation}
We note that in Eq. (\ref{eq:int2d}), the integration limits can be given as $a = -b = \req$, and $c = -d = y_{{\rm limit},i}$ for every $x_i$ coordinate.

The planetary spin axis is not necessarily perpendicular to the orbital plane. In such a case, two angles are introduced: the inclination of the spin axis, measured in the line of sight, and the obliquity ($\vartheta$), measured in the sky plane. The sky-projected coordinates ($X_{ij}$ and $Y_{ij}$) of the ellipse at time $t$ can therefore be calculated via a rotational matrix as
\begin{align}
    X_{ij} &= \delta_x(t) + x_{ij} \cos \left( \vartheta \right) - y_{ij} \sin \left( \vartheta \right) \\
    Y_{ij} &= \delta_y(t) + x_{ij} \sin \left( \vartheta \right) + y_{ij} \cos \left( \vartheta \right),
\end{align}
where $\delta_x(t)$ and $\delta_y(t)$ represent the position of the planetary center with respect to the center of the apparent stellar disk, normalized with the stellar radius $R_\star$. We show examples of the $X_{ij}$, $Y_{ij}$ points in Fig. \ref{fig:gl_transit} for a particular configuration.

In the general case, the planetary orbit is eccentric. Let $a/R_\star$, $i$, and $e$ denote the scaled semi-major axis, the inclination (with respect to the line of sight), and the eccentricity of the planetary orbit, and let $v(t)$ and $\omega$ denote its true anomaly and argument at the pericenter. Following \citep{2020MNRAS.496.4442C}, the distance between the sky-projected centers of the planet and the star, $\delta (t)$, is given by
\begin{equation}
\begin{split}
    \delta(t) & = \sqrt{\delta_x^2 + \delta_y^2} \\ &= \frac{a}{R_\star} \frac{1-e^2}{1+e \cos \left( v (t) \right)} \sqrt{1-\left(\sin \left(i\right)\right)^2 \left( \sin \left( v(t) + \omega \right) \right)^2}.
\end{split}    
\end{equation}

We calculate $X_{ij}$, $Y_{ij}$ for all integration points. The distance of each $X_{ij}$, $Y_{ij}$ point from the stellar center (assuming a spherical star) in normalized coordinates can be expressed as
\begin{equation}
    \mu = \sqrt{1-X_{ij}^2 - Y_{ij}^2}.
\end{equation}
When $\mu \notin \mathbb{R}$  (i.e. $X_{ij}^2 + Y_{ij}^2 > 1$), the integration point is outside the stellar disk (Fig. \ref{fig:gl_transit}). The stellar surface brightness ($I_{ij}$) at the $X_{ij}$, $Y_{ij}$ sky-projected coordinates is set to $0$ when $\mu$ is not finite. The darkening of the stellar limb is expressed in terms of the viewing angle $\gamma$, measured between the line of sight and the normal of the stellar surface, as $\cos \gamma = \mu$. When $\mu$ is finite (i.e. there is overlap between a point $X_{ij}$, $Y_{ij}$ within the boundary ellipse and the stellar contour), the planet is blocking light from the stellar surface. The amount of light blocked for any given point is determined by the darkening of the star's limb. In the Transit and Light Curve Modeller \citep[TLCM;][]{2020MNRAS.496.4442C}, there are seven different limb darkening laws, which can be used in these calculations. There is also an option to calculate without limb darkening. At the point $X_{ij}$, $Y_{ij}$, the stellar surface brightness depends on the chosen limb darkening law -- the possibilities are listed in Table \ref{tab:LD_laws}.

The viewing angle (or more precisely $\mu$) is calculated only when $\delta^2 < \left(1.1 \cdot \left( 1 + \req \right) \right)^2$. The numerical factor $1.1$ is arbitrarily selected -- any number $\geq 1$ would be acceptable. It's purpose is to provide a conservative estimate for the phases when a transit does occur. 

The flux loss during the transit of an oblate planet can be calculated as 
\begin{equation}
    \Delta F = \frac{2 \req}{4 \pi} \sum_j \sum_i w_j w_i y_{\rm{limit},i} I_{ij}.
\end{equation}
Out-of-transit, the stellar flux is given by
\begin{equation}
    F_{\star, {\rm out-of-transit}} = \int \int L_D \left( \mu \right) d\mu d \phi,
\end{equation}
where $\phi$ is the azimuthal angle, and $L_D$ is the surface brightness at the point ($\mu$, $\phi$) including the effect of limb darkening. The normalized transit light curve is given by
\begin{equation}
    F_{\rm transit} = \frac{\Delta F}{F_{\star, {\rm out-of-transit}}}.
\end{equation}
To speed up the calculations, $F_{\star, {\rm out-of-transit}}$ is computed once for the selected limb-darkening law. The expressions for the normalized transit light curve are given in Table \ref{tab:LD_laws}. We only utilized the quadratic limb darkening law in this work.

\begin{table}
    \centering
    \caption{Stellar surface brightness expressed as a function of various limb darkening laws implemented in TLCM and the corresponding transit of an oblate planet. The limb darkening coefficients are denoted by $u_{\rm lin}$, $u_{\rm a}$, $u_{\rm b}$, $C$, $\alpha$, $u_{\rm log, 1/2}$, $c_j$, $s_j$, and $r_{1/2}$.}
    \label{tab:LD_laws}
    \begin{tabular}{c c c c}
       \hline
       Name of law & $I_{ij}$ & Reference & $F_{\rm transit}$ \\
       \hline
       \hline
       No limb darkening & $1.0$ & -- & $1-\Delta F$ \\
       Linear &  $1 - u_{\rm lin} (1-\mu)$ & \tablenotemark{1} & $1 - \frac{\Delta F}{1-\frac{u_{\rm lin}}{3}}$ \\ 
       Quadratic & $1 - u_{\rm a} (1-\mu) - u_{\rm b} (1-\mu)^2$ & \tablenotemark{2} & $1 - \frac{\Delta F}{1-\frac{u_{\rm a}}{3}-\frac{u_{\rm b}}{6}}$ \\
       Power-2 & $1-C (1-\mu^\alpha)$ & \tablenotemark{3} & $1 - \frac{\Delta F}{1-\frac{C \alpha}{2+\alpha}}$ \\
       Logarithmic & $1-u_{\rm log, 1} (1 - \mu) - u_{\rm log, 2} \mu \ln \mu$ & \tablenotemark{4} & $1 - \frac{\Delta F}{1-\frac{u_{\rm log, 1}}{3}-\frac{2u_{\rm log, 2}}{9}}$ \\
       Claret-type four parameter law & $1- \sum_{j = 1}^4c_j (1-\mu^{j/2})$ & \tablenotemark{5} & $1 - \frac{\Delta F}{1-\frac{c_1}{5}-\frac{c_2}{3}-\frac{3c_3}{7}-\frac{c_4}{2}}$ \\
       Sing's law & $1- \sum_{j=2}^4 s_j (1-\mu^{j/2})$ & \tablenotemark{6} & $1 - \frac{\Delta F}{1-\frac{s_2}{3}-\frac{3s_3}{7}-\frac{s_4}{2}}$ \\
       Square-root & $1 - r_1 (1-\mu) - r_2 (1-\sqrt{\mu})$ & \tablenotemark{7} & $1 - \frac{\Delta F}{1-\frac{r_1}{3}-\frac{r_2}{5}}$ \\ 
       \hline
    \end{tabular}
    \tablenotetext{1}{\cite{1921MNRAS..81..361M}}
    \tablenotetext{2}{\cite{1985A&AS...60..471W}}
    \tablenotetext{3}{\cite{1997A&A...327..199H}}
    \tablenotetext{4}{\cite{1970AJ.....75..175K}}
    \tablenotetext{5}{\cite{2004A&A...428.1001C}}
    \tablenotetext{6}{\cite{2010A&A...510A..21S}}
    \tablenotetext{7}{\cite{1992A&A...259..227D}}
\end{table}

\begin{table}
    \centering
    \caption{Transit parameters used for simulating transits of spherical planets.}
    \label{tab:transitparams}
    \begin{tabular}{c c c c c c c}
        \hline         
       Case & $a/R_\star$ & $P$ [days] & $u_a$ & $u_b$ & $b$&$\rp$\\
        \hline
        \hline   
        \vspace{-.3 cm}
        & & & & \\        
      I &  73.29 & 87.0 & 0.2924 & 0.2924 & 0.25 &$\left\{ 0.04, 0.05, \dots, 0.20  \right\}$ \\
      II & 73.29 & 87.0 & 0.2924 & 0.2924 & $\left\{0, 0.05, \dots, 0.90 \right\}$ & 0.08 \\
      III & $\left\{5, 10, \dots, 100 \right\}$ & $\left\{ 1.55, \dots, 138.66 \right \}$ &   0.2924 & 0.2924 & 0.25 & 0.08\\
        \hline
    \end{tabular}
\end{table}

Additionally, the transits are characterized by the same parameters used for the modeling of spherical planets \citep{2020MNRAS.496.4442C}: semi-major axis ratio $a/R_\star$, the orbital period $P$, the time of midtransit $T_0$, and the conjunction parameter, described as
\begin{equation}
    b = \frac{a}{R_\star} \frac{\left(1-e^2\right) \cos \left( i \right)}{\left( 1 + e \sin \left( \omega \right) \right)},
\end{equation}
which simplifies to the well-known definition of the impact parameter for circular orbits.

\subsection{An improved model}

\begin{figure}
    \centering
    \includegraphics[width=\columnwidth]{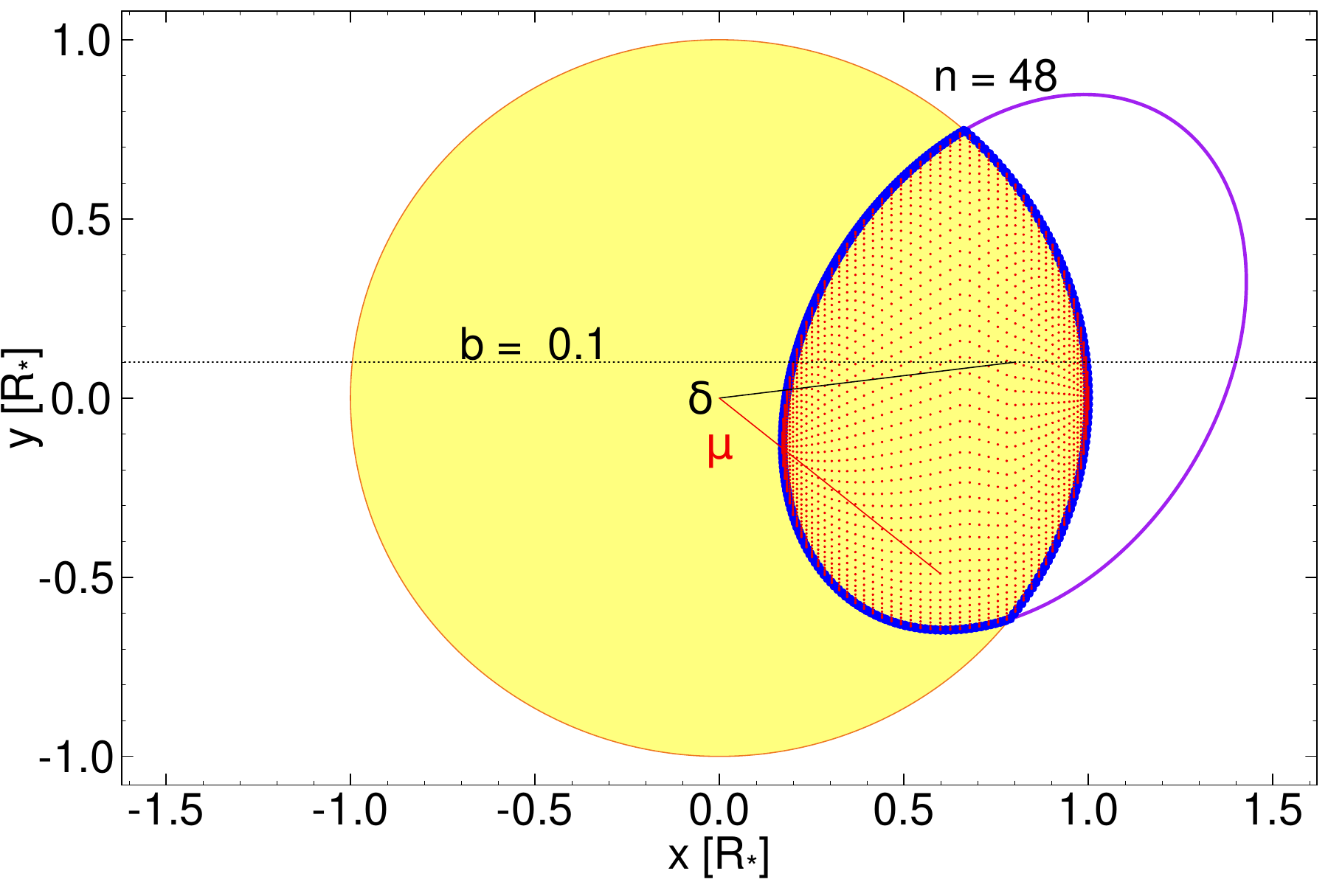}
    \caption{Dynamic distribution of the Gauss-Legendre quadrature points used for the transit light curve calculations of an oblate planet, from the improved model. The overlapping (transiting) area is highlighted in purple.}
    \label{fig:gl_transit_improved}
\end{figure}

Following the helpful suggestions of the referee after the initial submission, we also tested a revised and improved model for calculating the transit light curves. According to Eq. \ref{eq:glq}, the integration points ($x_i$) should cover the entire interval over which the integration is carried out. Consequently, the approach presented Sect. \ref{sec:glq} does not strictly follow the basic premises of this integration technique. In our revised approach, we first calculate the contour of the obscuring part of the planet (which consists of two arcs for the ingress/egress). We then distribute the integration points within these newly estimated boundaries dynamically, so the positions change according to $\delta(t)$. To do this, we divide the the contour of the planetary disk into evenly spaced units (in polar coordinates), then the $\vec{x}_i$ integration points can be calculated by interpolating between the nearest contour points. The $y_{{\rm limit},i}$ points can be computed by utilizing at the $\vec{x}_i$ points from the contours as well. The model is demonstrated in Fig. \ref{fig:gl_transit_improved}. We found that considerable improvements can be achieved by this dynamic distribution of the quadrature points, however, this comes at the cost of a runtime increase by two orders of magnitude (on the same computers). Given that the light curve modelings that are described below last  for $\approx$ 1 day, this is not feasible yet. A possible solution might be the utilization of GPU-based calculations, however, that is beyond the scope of this work.

\section{Performance of the model} \label{sec:tests}
\subsection{Limit of circular planet}
\begin{figure}
    \centering
    \includegraphics[width=0.5\columnwidth]{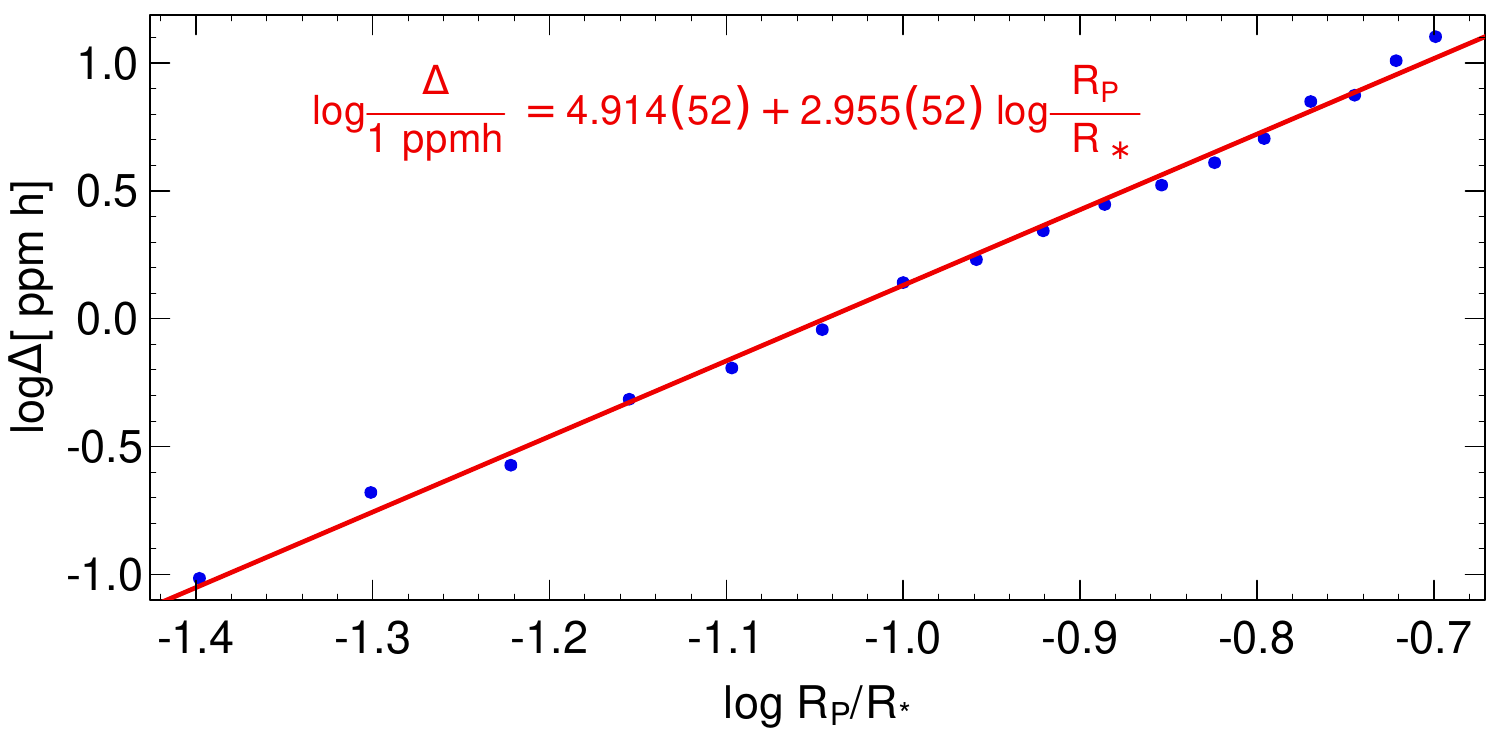}
    \caption{Discrepancy between the analytical and numerical models for a given set of transit parameters, with changing $\rp/R_\star$.}
    \label{fig:rp_diff}
\end{figure}

\begin{figure}
    \centering
    \includegraphics[width=0.5\columnwidth]{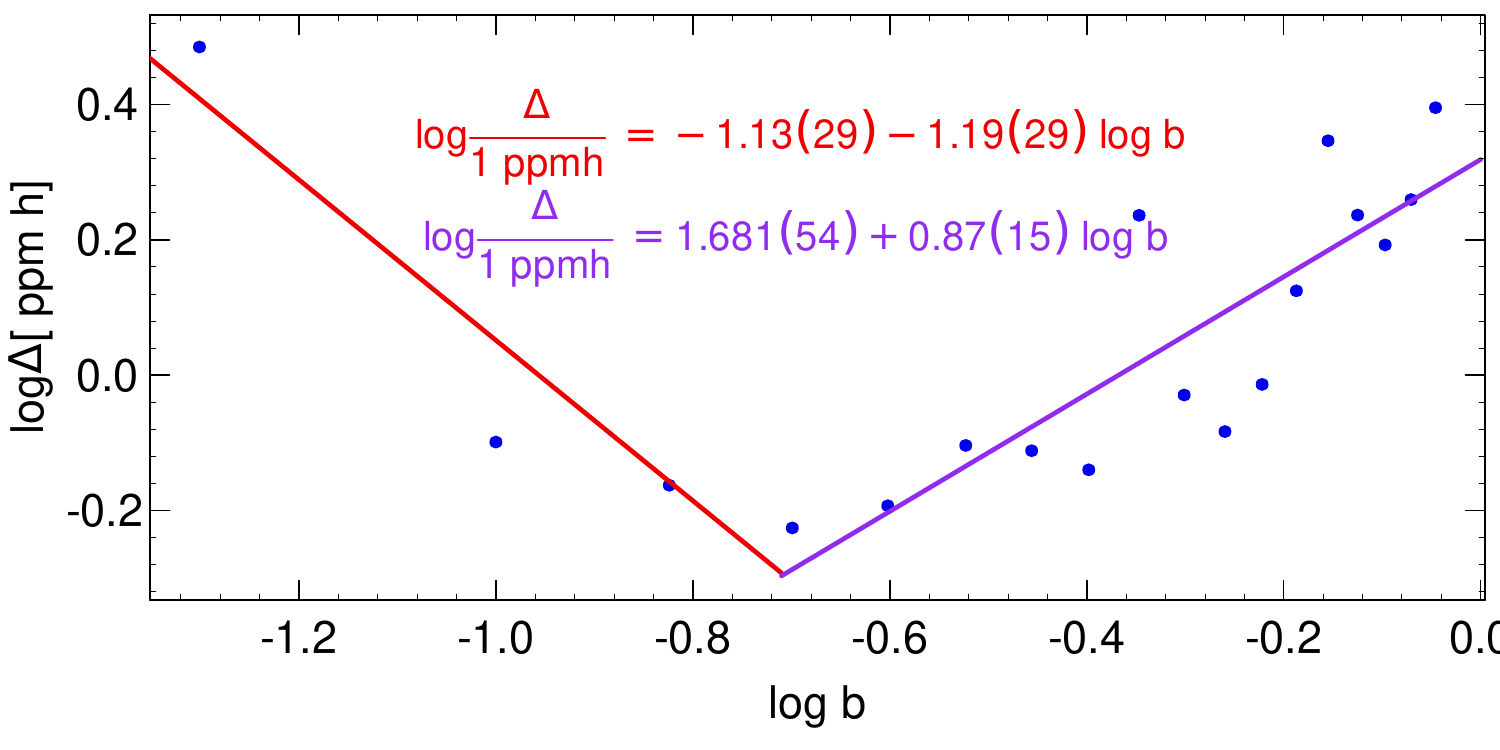}
    \caption{Discrepancy between the analytical and numerical models for a given set of transit parameters, with changing $b$.}
    \label{fig:b_diff}
\end{figure}

\begin{figure}
    \centering
    \includegraphics[width=0.5\columnwidth]{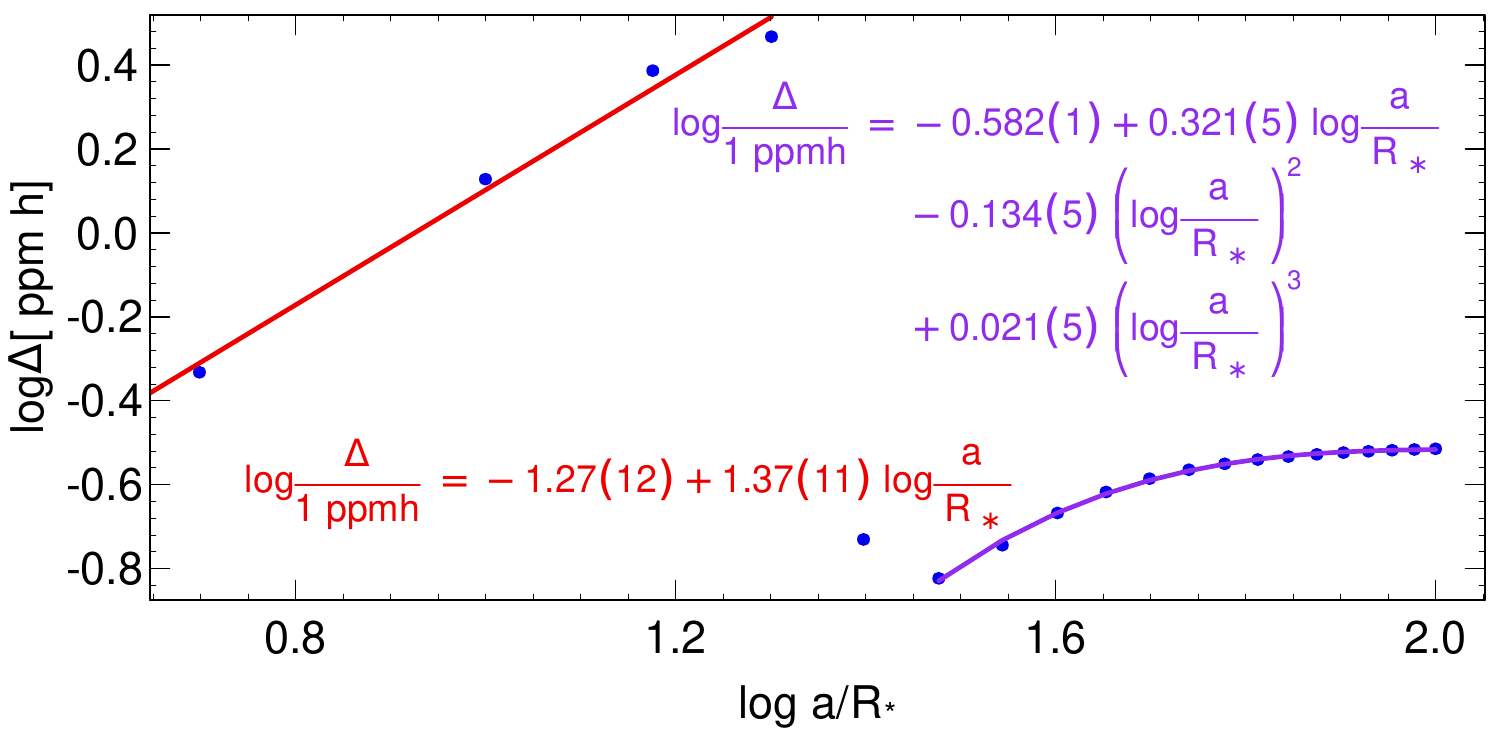}
    \caption{Discrepancy between the analytical and numerical models for a given set of transit parameters, with changing $a/R_\star$ (and corresponding $P$).}
    \label{fig:a_diff}
\end{figure}

\begin{table}
\centering
\caption{Values of the varied parameters and the corresponding $\Delta$ values in the three tested cases of the spherical planet.}
\label{tab:diffs}
\begin{tabular}{c c c c c c c }
\hline
\multicolumn{2}{c}{Case I} & \multicolumn{2}{c}{Case II} & \multicolumn{3}{c}{Case III} \\
$\rp/R_\star$ & $\log \Delta$ [ppm h] & $b$ & $\log \Delta$ [ppm h] & $a/R_\star$ & $P$ [days] & $\log \Delta$ [ppm h] \\
\hline
\hline
$0.04$ & $-1.02$ &  $0.00$ &  $0.54$ &  $5$ &  $1.55$ &  $-0.33$ \\ 
$0.05$ & $-0.68$ &  $0.05$ &  $0.48$ &  $10$ &  $4.38$ &  $0.13$ \\ 
$0.06$ & $-0.57$ &  $0.10$ &  $-0.10$ &  $15$ &  $8.06$ &  $0.39$ \\ 
$0.07$ & $-0.31$ &  $0.15$ &  $-0.16$ &  $20$ &  $12.40$ &  $0.47$ \\ 
$0.08$ & $-0.19$ &  $0.20$ &  $-0.23$ &  $25$ &  $17.33$ &  $-0.73$ \\ 
$0.09$ & $-0.04$ &  $0.25$ &  $-0.19$ &  $30$ &  $22.78$ &  $-0.82$ \\ 
$0.10$ & $0.14$ &  $0.30$ &  $-0.10$ &  $35$ &  $28.71$ &  $-0.74$ \\ 
$0.11$ & $0.23$ &  $0.35$ &  $-0.11$ &  $40$ &  $35.08$ &  $-0.67$ \\ 
$0.12$ & $0.34$ &  $0.40$ &  $-0.14$ &  $45$ &  $41.86$ &  $-0.62$ \\ 
$0.13$ & $0.45$ &  $0.45$ &  $0.24$ &  $50$ &  $49.02$ &  $-0.59$ \\ 
$0.14$ & $0.52$ &  $0.50$ &  $-0.03$ &  $55$ &  $56.56$ &  $-0.56$ \\ 
$0.15$ & $0.61$ &  $0.55$ &  $-0.08$ &  $60$ &  $64.44$ &  $-0.55$ \\ 
$0.16$ & $0.70$ &  $0.60$ &  $-0.01$ &  $65$ &  $72.66$ &  $-0.54$ \\ 
$0.17$ & $0.85$ &  $0.65$ &  $0.12$ &  $70$ &  $81.21$ &  $-0.53$ \\ 
$0.18$ & $0.87$ &  $0.70$ &  $0.35$ &  $75$ &  $90.06$ &  $-0.53$ \\ 
$0.19$ & $1.01$ &  $0.75$ &  $0.24$ &  $80$ &  $99.22$ &  $-0.52$ \\ 
$0.20$ & $1.10$ &  $0.80$ &  $0.19$ &  $85$ &  $108.66$ &  $-0.52$ \\ 
-- & -- &  $0.85$ &  $0.26$ &  $90$ &  $118.39$ &  $-0.52$ \\ 
-- & -- &  $0.90$ &  $0.40$ &  $95$ &  $128.39$ &  $-0.52$ \\ 
-- & -- &  -- &  -- &  $100$ &  $138.66$ &  $-0.51$ \\ 
\hline
\end{tabular}
\end{table}

In order to assess the intrinsic limitations of the approach described in Sect. \ref{sect:transit_calc}, a good baseline is needed. In the circular limit, when $f = 0$ (and thus $\req = \rpol = \rp$), the light curve should match the output of the well-established analytical methods introduced by \cite{2002ApJ...580L.171M}. Let us denote the light curve generated via the numerical calculations described above as $\phi_{\rm num} \left(t_j \right)$, and the analytical light curve as $\phi_{\rm MA} \left(t_j\right)$, both sampled at the same $t_j$ (discreet) time stamps. In order to quantify the discrepancies of the numerical model, we introduce the quantity $\Delta$ as the ``area under the curve'' of the residuals $\rho \left( t_j \right) = \phi_{\rm MA} \left( t_j \right) - \phi_{\rm num} \left( t_j \right)$:
\begin{equation} \label{eq:int_delta}
    \Delta = \sum_{k = 1}^{N-1} \left(t_{k+1}-t_k \right)\frac{\left(\rho \left(t_k \right) + \rho \left(t_{k+1} \right)  \right)}{2} .
\end{equation}
The definition of $\Delta$ is convenient even when dealing with unevenly sampled data.
We note that supersampling the residuals may be needed (depending on the cadence ($t_{j+1} - t_j$) of the light curve), and thus the total number of points $N$ on which $\Delta$ is calculated may exceed the total number of generated light curve points. This can be achieved via a linear interpolation. The quantity $\Delta$ (measured in, e.g., ppm $\times$ h) may be a useful tool to quantify the discrepancies between any two light curve models more generally than the case presented here. 

We generated light curves with both models using the parameters listed in Table \ref{tab:transitparams}. To assess how each of these parameters affects the precision of the numerical model, we conducted three experiments (cases I--III in Table \ref{tab:transitparams}). In these, we created a grid of $\rp/R_\star$ values (case I), a grid of $b$ values (case II) and a grid of $a/R_\star$ (and consequently corresponding $P$ values, case III). In each case, we varied only one of the three basic transit parameters and measured $\Delta$ at each grid point. As the transits are symmetrical, the time of midtransit (the fourth basic transit parameter) does not affect the discrepancy between the two models. The limb darkening coefficients can take on arbitrary values depending on the star and the observing bandpass, we therefore elected to keep these the same throughout the tests. As a consequence of the gravitational interaction between the two bodies (the star and its companion), the semi-major axis and the orbital period are not independent parameters. Assuming the same $\massp$ in every case, we can find the corresponding $P$ to every $a/R_\star$ using Kepler's third law:
\begin{equation}
    \frac{\left(\frac{a}{R_\star}\right)^3}{P^2} = \frac{73.29^3}{87^2 {\rm d}^2}
\end{equation}
and with parameters taken from Table \ref{tab:transitparams}.

\begin{figure}
    \centering
    \includegraphics[width = \columnwidth]{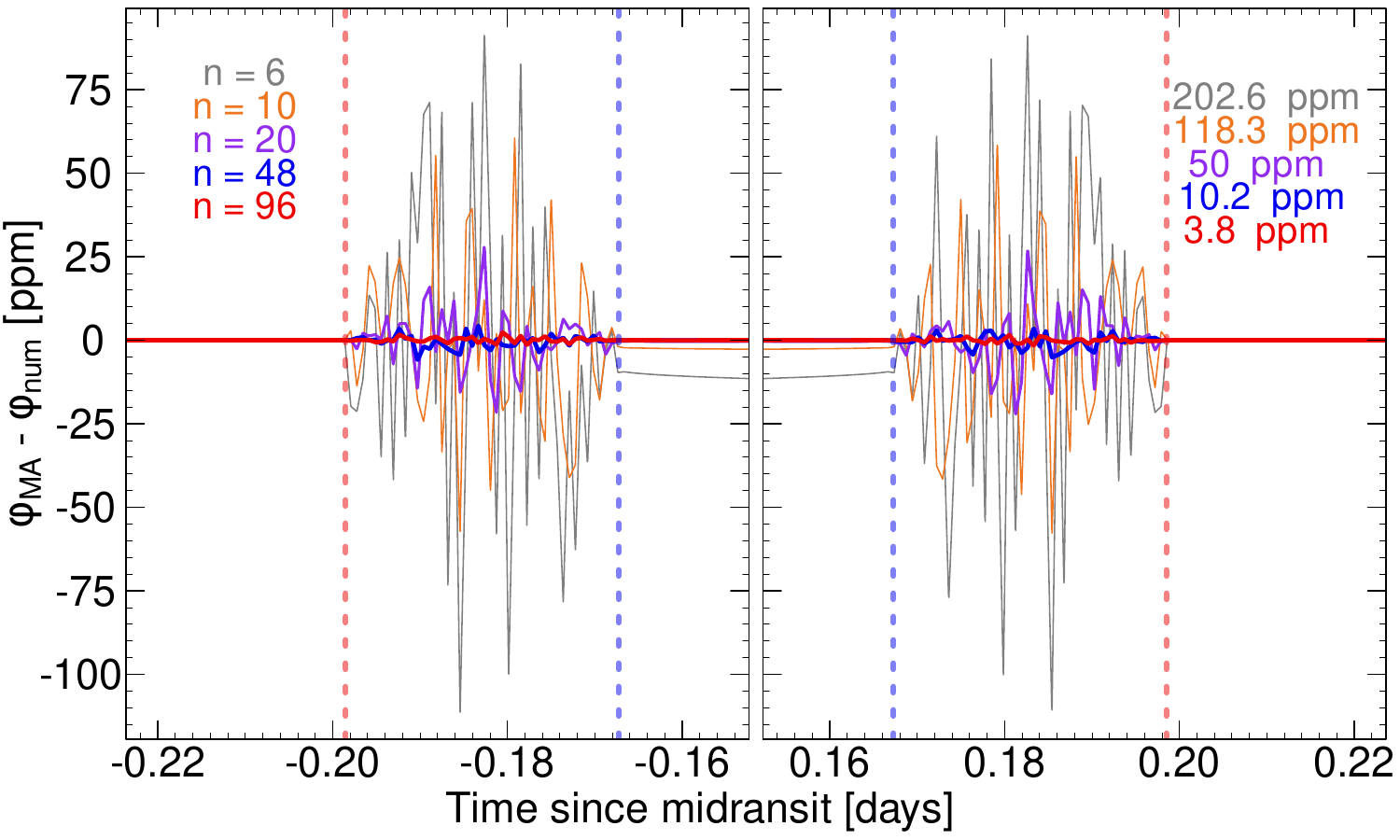}
    \caption{Difference between the analytical model ($\phi_{\rm MA}$) and the numerical model presented here ($\phi_{\rm MA}$) for a transit generated with parameters taken from case I (Table \ref{tab:transitparams}) with $\rp/R_\star = 0.08$, and $n \in \{ 6, 10, 20, 48, 96 \}$ GL quadrature points. The dashed red lines mark the first and fourth contact points, the dashed blue lines mark the second and third contact points. The peak-to-peak amplitudes of the residuals are shown with grey (for $n = 6$), orange ($n = 10$), purple ($n = 20$), blue ($n = 48$), and red ($n = 96$).}
    \label{fig:residuals}
\end{figure}

We calculated the transits on a grid of $17$ $\rp$ values, $19$ different impact parameters, and $20$ semi-major axes. We used $n = 96$ GL quadrature points for every simulation. The resultant $\Delta$ values are listed in Table \ref{tab:diffs}, and are shown in Figs. \ref{fig:rp_diff}, \ref{fig:b_diff}, and \ref{fig:a_diff}. The dominant factor in the discrepancy between the analytical and numerical transit calculations is the size of the planet. This is because the same $n$ number of GL quadrature points is used to cover a larger surface of the (limb-darkened) stellar disk, yielding lower precision during the partial transits (between the first and second contact points, as well as the third and fourth contact points, Fig. \ref{fig:residuals}). We find a cubic relationship between $\Delta$ and $\rp/R_\star$:
\begin{equation}
    \log \left( \frac{\Delta}{1~{\rm ppm\;h}}\right) = -2.914(52) \pm 2.955(52) \log \left( \frac{\rp}{R_\star} \right),
\end{equation}
as established via a least squares fit (\ref{fig:rp_diff}). The observed $\Delta$ values in case I range over more than two orders of magnitudes, from $9.66  \cdot 10^{-2}$~ppm h (at $\rp = 0.04 R_\star$) to $12.7$~ppm h (at $\rp = 0.20 R_\star$). As a direct point of comparison, the residuals shown in Fig. \ref{fig:residuals} show a peak-to-peak  amplitude of $\approx 3.3$ ppm, which in this case corresponds to $\Delta = 0.65$~ppm h. The latter quantity, however, is a better representation of the error as it also contains information about how many timestamps have non-zero residuals. 

The other explored parameters have a comparably lower effect on $\Delta$, as in both cases II and III it only spans over one order of magnitude (Table \ref{tab:diffs}). The $\Delta$ --$b$ and $\Delta$--$a/R_\star$ relationships (Figs. \ref{fig:b_diff} and \ref{fig:a_diff}) are not as easy to describe as with the planet size, however. There appears to be a minimum in $\Delta$ at $b \approx 0.2$ ($b = 0$ can not be shown on a logarithmic scale), with $\Delta = 0.58$~ppm h. We find
\begin{equation}
        \log \left( \frac{\Delta}{1~{\rm ppm\;h}}\right) =
        \begin{cases}
            -1.13(29) - 1.19(29)\log b & \text{if $b \lessapprox 0.2$} \\
            1.681(54) + 0.87(15)\; \log b & \text{if $b \gtrapprox 0.2$}.\\
        \end{cases}
\end{equation}
For the larger $b$ values, the uncertainty term increases linearly with the increase in $b$.

In the case of $a/R_\star$, we find two distinct patterns (Fig. \ref{fig:a_diff}) described by
\begin{equation}
    \log \left( \frac{\Delta}{1~{\rm ppm\;h}}\right) =
    \begin{cases}
        -1.27(12) + 1.37(11)\log \left( \frac{a}{R_\star} \right) & \text{if } a/R_\star \lessapprox 20, \\[8pt]
        -0.582(1) + 0.321(5)\log \left( \frac{a}{R_\star} \right) 
         - 0.314(5) \left(\log \left( \frac{a}{R_\star} \right) \right)^2 
        + 0.021(5) \left( \log \left( \frac{a}{R_\star} \right)\right)^3 & \text{if } a/R_\star  \gtrapprox 20.
    \end{cases}
\end{equation}
The impact parameter has an effect on the transit depth (because of the stellar limb darkening) and the transit duration. However, we do not observe a difference between the transit depths of $\phi_{\rm MA}$ and $\phi_{\rm num}$ (Fig. \ref{fig:residuals}). The parameters varied in cases II and III have an influence on the transit duration, and therefore, for a given exposure time ($60$ seconds throughout these tests), there are fewer light curve points during the ingress/egress of the transit at lower $a/R_\star$ and higher $b$. The variability in the $\rho$ residuals may increase during these phases as a consequence. On the other hand, at $b \approx 0$, the limb-darkened stellar surface will be sampled at positions where the surface brightness $L_D$ changes rapidly with $\mu$ (and consequently the sky-projected position of the planet) -- causing increased scatter in the residuals. For long-period planets ($a/R_\star \gtrsim 50$), there is a monotonous decrease in the rate of increase of $\Delta$ with higher periods. This may be caused by the decreasing change in sky-projected planetary position at each time step, which in turn results in a more precise sampling of $L_D$ during the critical phases of the transit. A finer $a/R_\star$ ($P$) and $b$ grid may reveal deeper connections between the scatter of the residuals and the basic transit parameters, however, that is beyond the scope of this paper.

We note that the intrinsic limitation of the GL integration appears to be on the order of several ppm in amplitude. This could be reduced further by increasing the number of quadrature points; however, even the precision should be adequate for the detection of oblate planets, as is discussed below.


\subsection{Efficiency}
\begin{figure}
    \centering
\includegraphics[width=0.5\columnwidth]{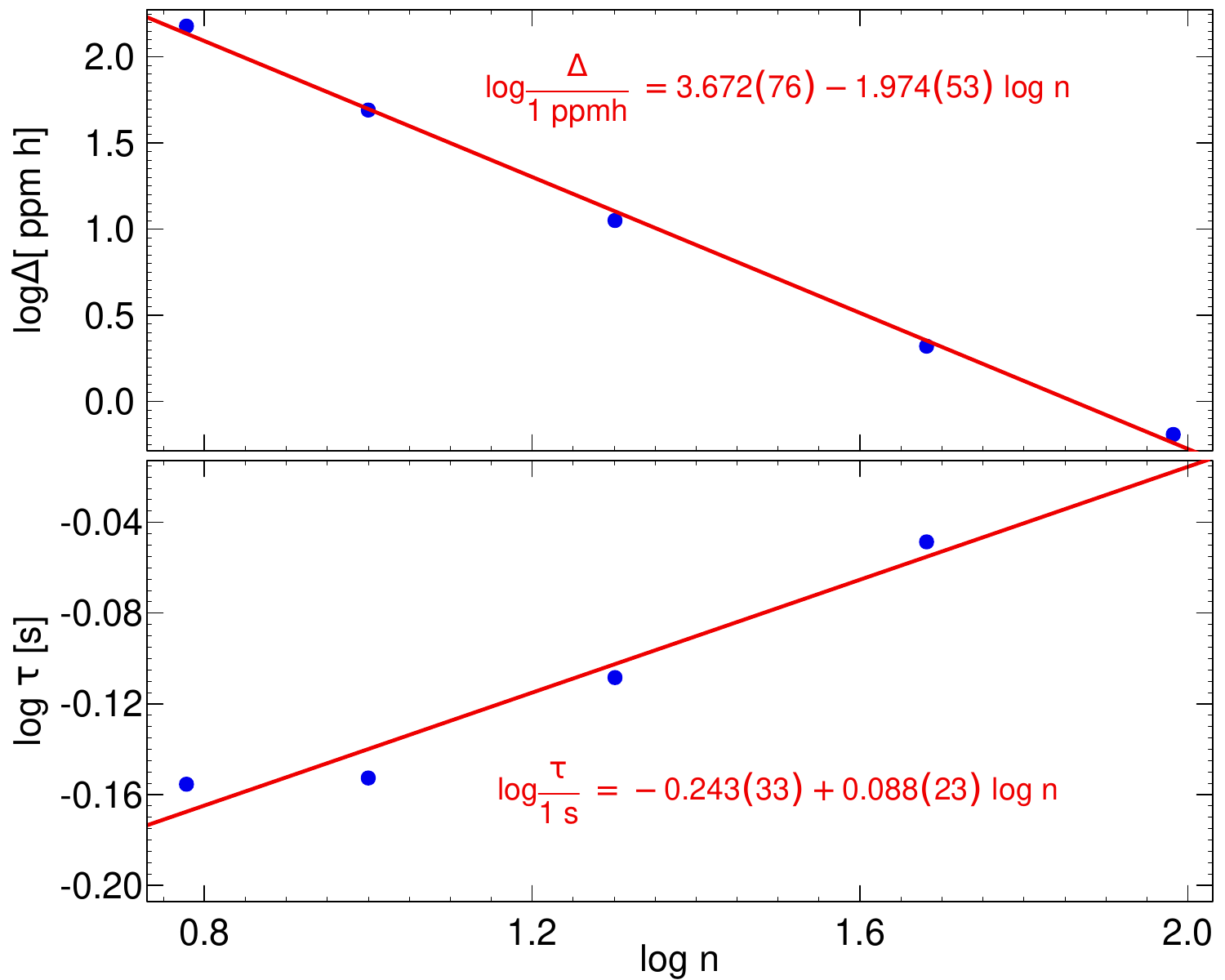}
    \caption{Discrepancy between the analytical and numerical models for a given set of transit parameters, with changing $n$ number of Gauss-Legendre quadrature points (top panel). The light curve calculation time is shown on the bottom panel.}
    \label{fig:effi}
\end{figure}
\begin{table}[!h]
\centering
\caption{Distortion signal ($\log \Delta$, expressed in ppm h) caused by the oblateness of the planet on the $f$ -- $\vartheta$ grid. }
\label{tab:snr}
\begin{tabular}{|c | c c c c c c c c c c|}
\hline
\diagbox{$\vartheta$ [$^\circ$]}{$f$} & $0.03$ & $0.06$ & $0.09$ & $0.12$ & $0.15$ & $0.18$ & $0.21$ & $0.24$ & $0.27$& $0.30$ \\
\hline
\hline
$0$ & $1.25$ &  $1.53$ &  $1.69$ &  $1.79$ &  $1.87$ &  $1.92$  &  $1.96$  &  $2.00$  &  $2.02$  &  $2.04$ \\ 
$9$ & $1.18$ &  $1.46$ &  $1.61$ &  $1.71$ &  $1.79$ &  $1.84$  &  $1.88$  &  $1.91$  &  $1.93$  &  $1.95$ \\ 
$18$ & $1.05$ &  $1.32$ &  $1.48$ &  $1.57$ &  $1.64$ &  $1.69$  &  $1.74$  &  $1.77$  &  $1.80$  &  $1.83$ \\ 
$27$ & $1.02$ &  $1.31$ &  $1.47$ &  $1.58$ &  $1.66$ &  $1.73$  &  $1.78$  &  $1.82$  &  $1.85$  &  $1.88$ \\ 
$36$ & $1.03$ &  $1.32$ &  $1.48$ &  $1.59$ &  $1.68$ &  $1.74$  &  $1.79$  &  $1.83$  &  $1.86$  &  $1.88$ \\ 
$45$ & $1.03$ &  $1.30$ &  $1.46$ &  $1.57$ &  $1.64$ &  $1.71$  &  $1.75$  &  $1.79$  &  $1.82$  &  $1.84$ \\ 
$54$ & $1.07$ &  $1.36$ &  $1.54$ &  $1.65$ &  $1.75$ &  $1.82$  &  $1.88$  &  $1.93$  &  $1.98$  &  $2.02$ \\ 
$63$ & $1.20$ &  $1.49$ &  $1.66$ &  $1.78$ &  $1.87$ &  $1.94$  &  $1.99$  &  $2.04$  &  $2.08$  &  $2.12$ \\ 
$72$ & $1.27$ &  $1.56$ &  $1.73$ &  $1.84$ &  $1.93$ &  $2.00$  &  $2.05$  &  $2.10$  &  $2.14$  &  $2.17$ \\ 
$81$ & $1.29$ &  $1.58$ &  $1.75$ &  $1.86$ &  $1.95$ &  $2.02$  &  $2.07$  &  $2.12$  &  $2.16$  &  $2.19$ \\ 
\hline
\end{tabular}
\end{table}
We test the efficiency of the model by comparing $\phi_{\rm num}$ light curves generated with the full range of possible GL integration points to $\phi_{\rm MA}$. The residuals for $n \in \{6, 10, 20, 48, 96\}$, taken from case I (Table \ref{tab:transitparams}) with $\rp/R_\star = 0.08$ are shown in Fig. \ref{fig:residuals}. At $n = 6$ and $n = 10$, the numerical approach yields transits that are $\approx 12$ and $\approx 3$ ppm deeper than the baseline analytical model. Above $n = 20$, the transit depths are in excellent agreement between $\phi_{\rm num}$ and $\phi_{\rm MA}$. The discrepancies arise during the partial transits (i.e. between the first and second, and third and fourth contacts). The increased number of integration points decreases the peak-to-peak amplitude of the ``scatter'' in the light curves from $\approx 203$ ppm to $\approx 4$ ppm (when comparing $n = 6$ with $n = 96$). The corresponding $\Delta$ values also decrease by $\gtrsim 2$ orders of magnitude (Table \ref{tab:effi}) from $\approx 151$ ppm h to $\approx 0.65$ ppm h. We find that the discrepancy between the numerical and analytical models can be described as
\begin{equation}
    \log \left( \frac{\Delta}{1~{\rm ppm\;h}}\right) = 3.672(76) - 1.974(53) \log n,
\end{equation}
which is within $1 \sigma$ of a quadratic improvement with the increase in GL quadrature points (Fig. \ref{fig:effi}). 

We also measure the runtime $\tau$ of a light curve generation. We model one transit, with a $60$ s exposure time, and parameters taken from case I (Table \ref{tab:transitparams}, with $\rp/R_\star = 0.08$). We compute the light curves on an \verb|12th Gen Intel Core i9-12950HX| processor. The resultant runtimes are shown in Table \ref{tab:effi}. The total runtime increases with 
\begin{equation}
    \log \left( \frac{\Delta}{1~{\rm ppm\;h}}\right) = -0.243(33) + 0.088(23) \log n,
\end{equation}
which is a much weaker relationship than linear. For that reason, we use $n = 96$ quadrature points in the simulations below. We note that $n = 48$ and in some cases even $n = 20$ points could yield adequate results. We also note that $\tau$ for the analytical model is $1.142$ seconds on the same processor, which is $\approx 28\%$ more than even the $n = 96$ case.
\begin{table}
\centering
\caption{Computational time ($\tau$) and the discrepancy between the analytical and numerical models for the tested GL integration points.}
\label{tab:effi}
\begin{tabular}{c c c}
\hline
$n$ & $\tau$ [s] & $\log \Delta$ [ppm h] \\
\hline
\hline
$6$ & $0.696$ &  $2.18$\\ 
$10$ & $0.699$ &  $1.69$\\ 
$20$ & $0.704$ &  $1.05$\\ 
$48$ & $0.779$ &  $0.32$\\ 
$96$ & $0.894$ &  $-0.19$\\ 
\hline
\end{tabular}
\end{table}

\subsection{Comparison with squishyplanet}

\begin{figure}
    \centering
    \includegraphics[width=0.5\linewidth]{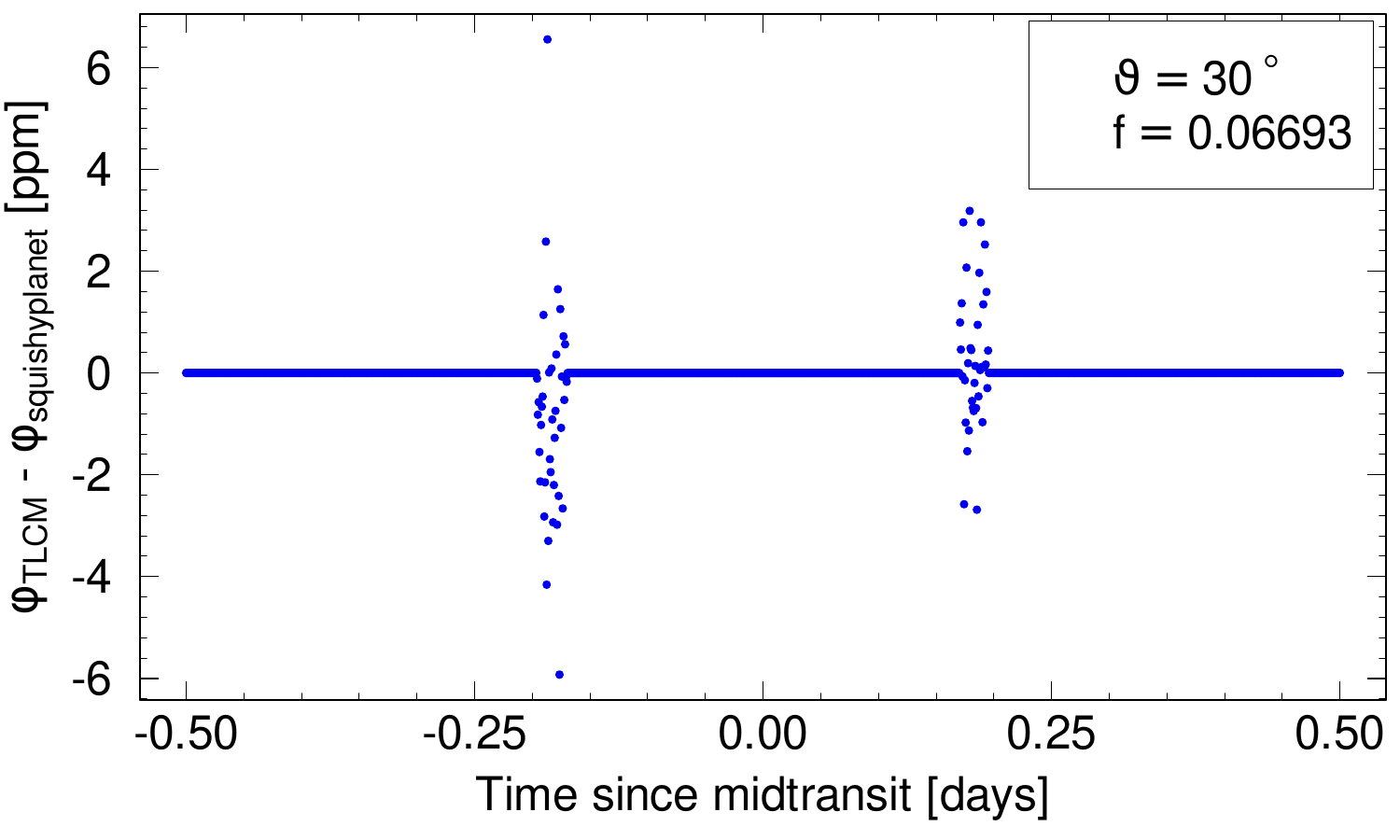}
    \caption{Difference between a transit light curve of an oblate planet simulated with TLCM and squishyplanet.}
    \label{fig:squishy}
\end{figure}

In order to assess the performance of our model with respect to other, analytical frameworks for simulating the transit light curves of oblate planets, we simulated a transit using squishyplanet \citep{2024JOSS....9.6972C} and TLCM. We used the following parameters: $a/R_\star = 73.29$, $P = 87$ days, $u_a = u_b = 0$, $b = 0.24$, $\req = 0.069289$, $f = 0.06693$, and $\vartheta = 30^\circ$. The difference between the two light curves is shown on Fig. \ref{fig:squishy}. The amplitude of the difference between the two light curves is $\approx 13$~ppm. Using the area of the sky-projected ellipse, we can calculate the effective radius of a spherical planet which would yield the same transit depth as $R_{\rm eff} = \rp \cdot \sqrt{1-f}$. By also calculating a transit light curve with $R_{\rm eff}$ with the Mandel-Agol model, we estimate the amplitude of the oblateness signal to be $\approx 45$~ppm in this case - so in theory a detection of $>3 \sigma$ is feasible with our model. By using Eq. (\ref{eq:int_delta}), we find that the error term, compared to squishyplanet is $ \Delta \approx 4.6$ ppm$\times$h, while for the oblateness signal $\Delta = 21$ ppm$\times$h, also implying the possibility of a $> 3 \sigma$ detection.

\cite{2003ApJ...588..545B} found that when modeling the light transit light curve of an oblate planet with a model for a spherical planet, a better fit is achieved by modifying the orbital parameters (primarily the impact parameter) as well as the planetary radius. This implies that the oblateness signal seen on Fig. \ref{fig:oblate_residuals} is optimistic -- in a real-world scenario without the a priori knowledge of the transit parameters it would likely be lowe leading to lower estimates on $\delta$).

\section{Results} \label{sec:results}
\begin{figure}
    \centering
    \includegraphics[width=\linewidth]{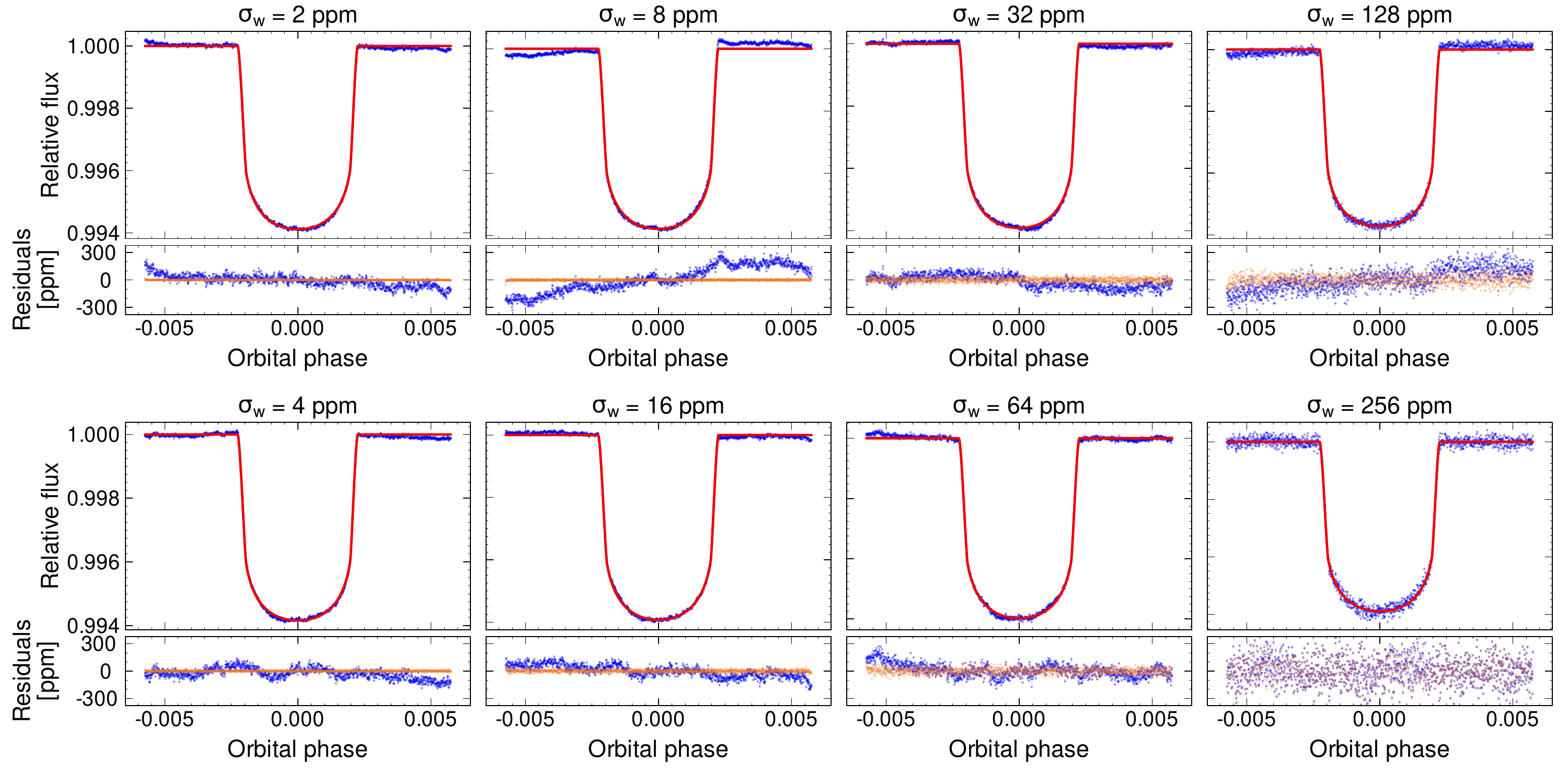}
    \caption{Example light curves with red noise (bigger panels, blue dots) for $\vartheta = 63^\circ$ and $f = 0.21$ at all 8 $\sigma_w$ noise levels. The best-fit transit models are shown with solid red lines. The residuals are shown on the smaller panels for every noise level. Orange dots show the residuals after the removal of the fitted red noise -- ideally these are characterized by the respective $\sigma_w$ standard deviations.}
    \label{fig:demo_lcs}
\end{figure}
We performed injection-and-retrieval tests on a large number of light curves with different noise levels and oblanteness-related parameters to assess the feasibility of detecting the non-spherical shape of transiting exoplanets. We analysed one transit per light curve.
\subsection{Parameter grid} \label{sec:grid}
\begin{figure}
    \centering
    \includegraphics[width=\linewidth]{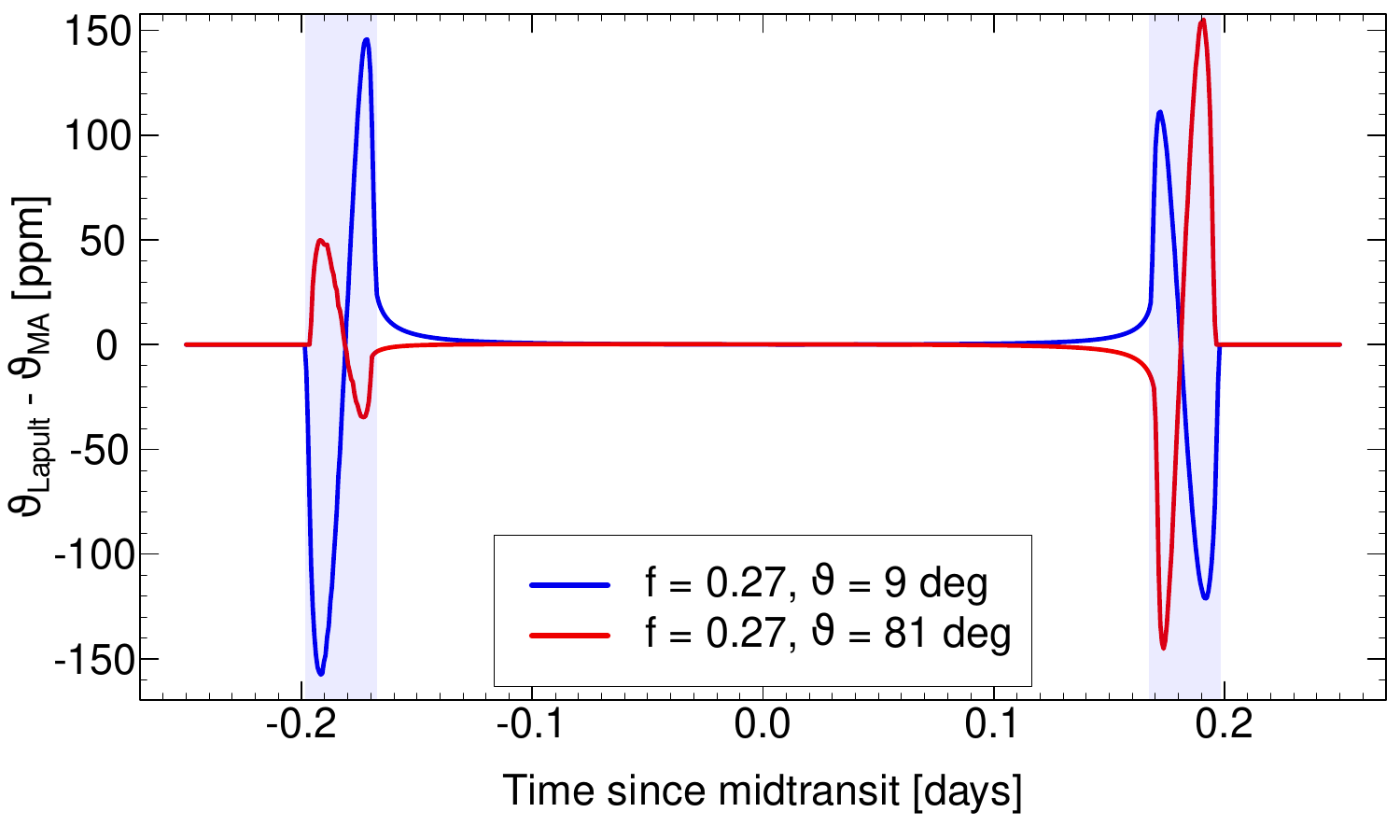}
    \caption{Difference between the transit light curve of an oblate planet and a spherical planet, for a particular choice of $f$ and $\vartheta$. The shaded regions highlight the ingress and egress phases of the transit.}
    \label{fig:oblate_residuals}
\end{figure}

\begin{figure}
    \centering
    \includegraphics[width=0.5\linewidth]{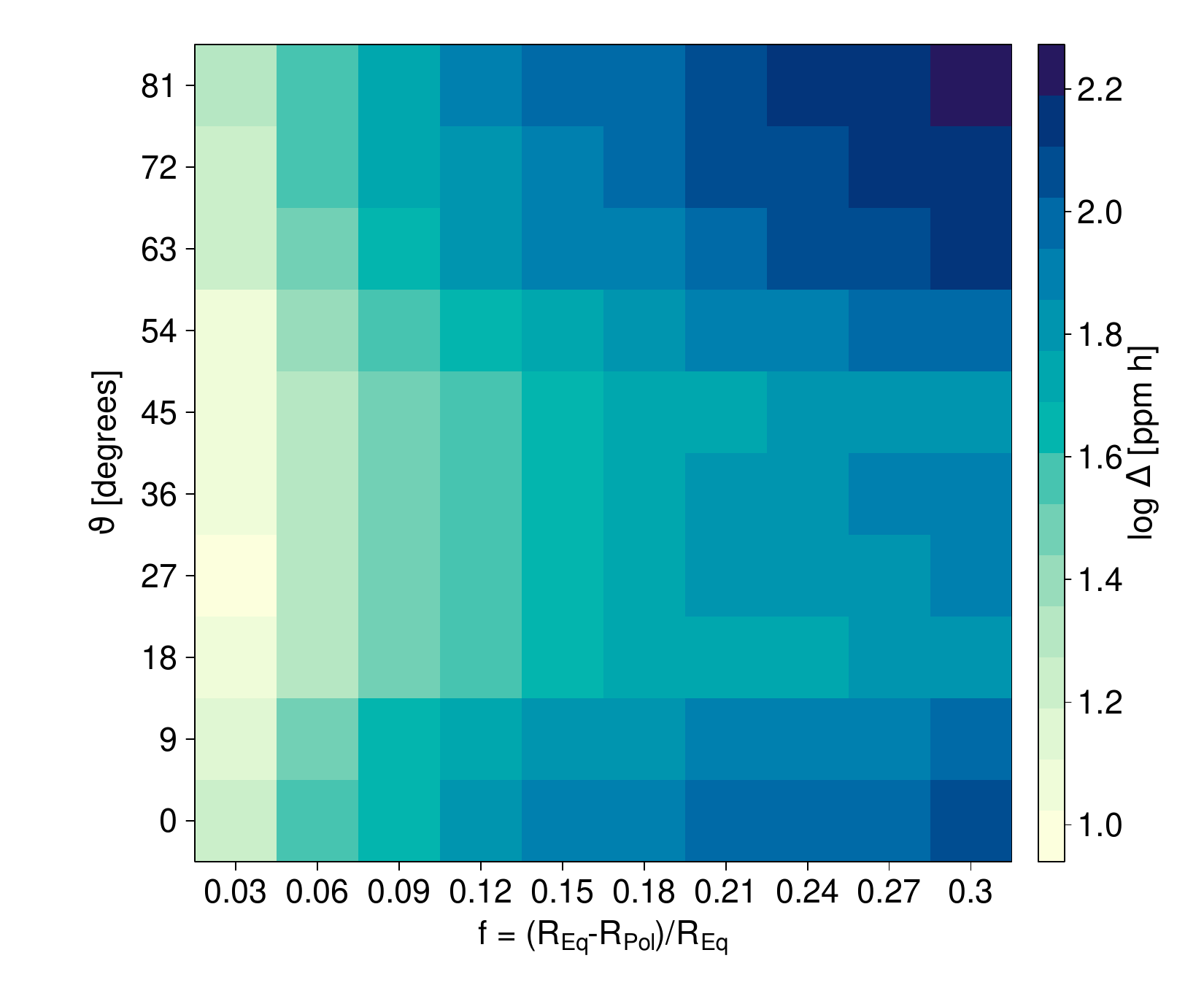}
    \caption{Distortion signal in the transit light curve cause by the oblateness at every $f$--$\vartheta$ grid point as compared to a circular planet. The distortion $\Delta$ is expressed as the area under the curve of the residuals of an oblate planet and a circular one with the same effective radius.}
    \label{fig:snr}
\end{figure}

We performed a large-scale test to assess the feasibility of detecting the oblateness of exoplanets with the the model described in \ref{sect:transit_calc}. We generated transits of oblate planets using the following transit parameters: $a/R_\star = 73.26$, $P = 87$ days, $b = 0.25$, and $\req = 0.08$. We set the limb darkening at $A = B = 1.3$, roughly corresponding to a Solar-like star using FGS2 (Fine Guidence Sensor) of \textit{ARIEL} We generated light curves on a grid of $f \in \{0.03, 0.06, \dots, 0.30\}$ and $\vartheta \in   \{0^\circ, 9^\circ, \dots, 90^\circ\}$. This process yielded $100$ different configurations. The transits were generated using $n = 96$ GL quadrature points. Let $\phi_{\rm oblate}$ denote the transit light curve of an oblate planet. Using the area of the sky-projected ellipse, we can calculate the effective radius of a spherical planet which would yield the same transit depth as $R_{\rm eff} = \rp \cdot \sqrt{1-f}$. Let $\phi_{\rm MA}$ denote the transit light curve of a spherical planet with the radius $R_{\rm eff}$. According to Eq. (\ref{eq:int_delta}), we can then measure the effect of oblateness on the transits from on the $\phi_{\rm oblate}-\phi_{\rm MA}$ residuals, as $\Delta$ will correspond to the distortion caused by the ellipsoidal shape of the planet (also incorporating the systematic noise of the model, as discussed in Sect. \ref{sec:tests}). The residuals for two configurations are shown in Fig. \ref{fig:oblate_residuals}. The measured $\Delta$ values on the $f$ -- $\vartheta$ grid are shown in Fig. \ref{fig:snr} and in Table \ref{tab:snr}.

We tested eight different white noise levels that are characterized by their $\sigma_w$ point-to-point scatter. We used a grid of $\sigma_w \in \{1, 2, 4, \dots, 256\}$ ppm, for the $60$ s exposure time used in the light curve simulations. As a final layer of the grid, we also solved these light curves in the presence of red noise and without it. The time-correlated noise model was chosen as the Autoregressive Integrated Moving Average (ARIMA) clone of a Hubble Space Telescope observation. The noise model is described in Section 3.2 of \cite{kalman24_agent}. We scaled the amplitude of the time-correlated noise as presented in \cite{kalman24_agent} by a factor of $1/10$. Each light curve is exactly one-day long, and is centered on the conjunction of the planet. We injected a randomly selected segment of the noise model into the transit light curve. Every one-day long noise model is chosen independently from the others.

We therefore performed $800$ light curve analyses. Example light curves for every $\sigma_w$ are shown on Fig. \ref{fig:demo_lcs}.

\subsection{Light curve fitting}

We analyzed the light curves using $n = 96$ GL integration points for every transit, utilizing TLCM. We made use of a DE-MCMC \citep[Differential Evolution Markov-chain Monte Carlo;][]{2006ApJ...642..505F} process of $10$ chains with $10,000$ steps each to explore the parameter space of the transit parameters. If convergence is not reached within the pre-defined number of steps, the chain lengths are automatically increased, up to ten times the original lengths. We identified a degeneracy between $b$, $f$, and $\vartheta$, confirming the findings of \cite{2024arXiv241003449D}. On the one hand, this degeneracy can increase the uncertainty range of the oblateness-related parameters (also possibly hiding a detection of the true planetary oblateness). On the other hand, it can also cause the MCMC algorithm to get stuck in a local minimum (of the negative log-likelihood), thus returning with an inaccurate set of parameters. 

There is no independent way of measuring the orbital inclination (and consequently the impact parameter). However, since both $b$ and $a/R_\star$ are related to the transit duration \citep{2003ApJ...585.1038S}, we found that by placing constraints on the semi-major axis, the degeneracies can be broken. By rearranging Kepler's third law, the semi-major axis can be expressed as \citep[e.g.][]{2020MNRAS.496.4442C}:
\begin{equation}
    \left( \frac{a}{R_\star} \right)^3 = \frac{P^2 G \left( 1 + q \right)}{3 \pi} \rho_\star,
\end{equation}
where $q = \massp/M_\star$ is the planet-to-star mass ration, and $\rho_\star$ is the stellar density. Assuming a well-characterized orbital period and neglecting the uncertainty in the mass of the planet, we can write that the relative uncertainty is:
\begin{equation} \label{eq:rho}
    \frac{\Delta \left(\frac{a}{R_\star} \right)}{\frac{a}{R_\star}} = \frac{1}{3} \frac{\Delta \rho_\star}{\rho_\star}.
\end{equation}
\cite{2017ApJ...835..173S} found that by utilizing asteroseismology based on \textit{Kepler} photometry \citep{2010Sci...327..977B}, the mean density of dwarf stars can be determined with a precision between $0.5\%$ and $2.6\%$. It is  expected that similar levels of  precision can be achieved with \textit{PLATO} \citep{2014ExA....38..249R,2024arXiv240605447R} as well. Consequently, we may assume $\Delta \rho_\star/\rho_\star = 0.01$, which translates to $\Delta \left(a/R_\star \right)/(a/R_\star) = 0.0033$ according to Eq. (\ref{eq:rho}). As a result, we applied a Gaussian prior on the scaled semi major axis with a mean of $73.26$ and a standard deviation of $0.24$. We also placed strict priors in the limb-darkening coefficients $A$ and $B$, with standard deviations of $0.01$ in both cases. We note that this trick is valid strictly for circular orbits. In a more general case, the orbital eccentricity also has an effect on the transit duration \citep[see e.g.][]{2012ApJ...756..122D}, thus it likely further complicates the parameters recovery. We suggest that the orbital eccentricity (and the argument of the periastron) also need to be known a priori, from independent observations (e.g. radial velocity measurements), to facilitate the parameter recovery shown here.

The time-correlated noise is modeled using the wavelet-based algorithm of \citep{2009ApJ...704...51C}. It is described by two parameters: $\sigma_r$ for the red noise and $\sigma_w$ for the white noise. In TLCM, there is a penalty function which ensures that the average photometric uncertainty must be equal to $\sigma_w$, to avoid overfitting \citep{2023A&A...675A.106C}. In addition, we also fitted for a constant offset, $p_0$. The list of parameters used in the analysis, as well as their priors, are listed in Table \ref{tab:parameter_setup}.

\subsection{Parameter recovery} \label{sec:param_recovery}

\begin{figure}
    \centering
    \includegraphics[width=\linewidth]{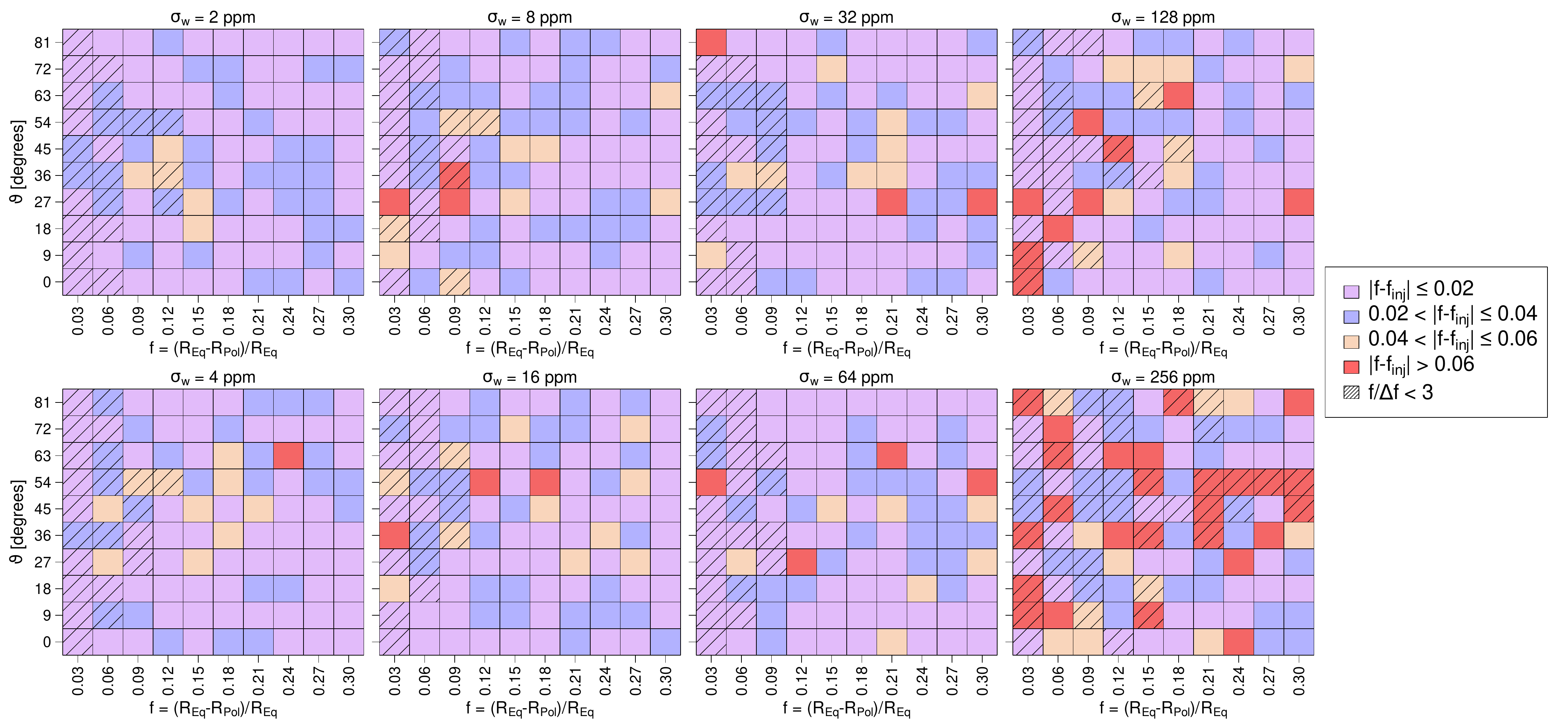}
    \caption{Accuracy of the retrieved oblateness parameter in the light curve which included red noise, for all eight $\sigma_w$ noise levels. Squares in the input $f$--$\vartheta$ grid are coloured based on the accuracy of the fitted $f$: purple if the retrieved parameter is within $0.02$ of the injected value, blue if it is within $0.04$, orange when it is within $0.06$, and red otherwise. Shading is present when no significant detection of oblateness is made (i.e. when there is no $3 \sigma$ detection of $f$).}
    \label{fig:f_abs}
\end{figure}

\begin{figure}
    \centering
    \includegraphics[width=\linewidth]{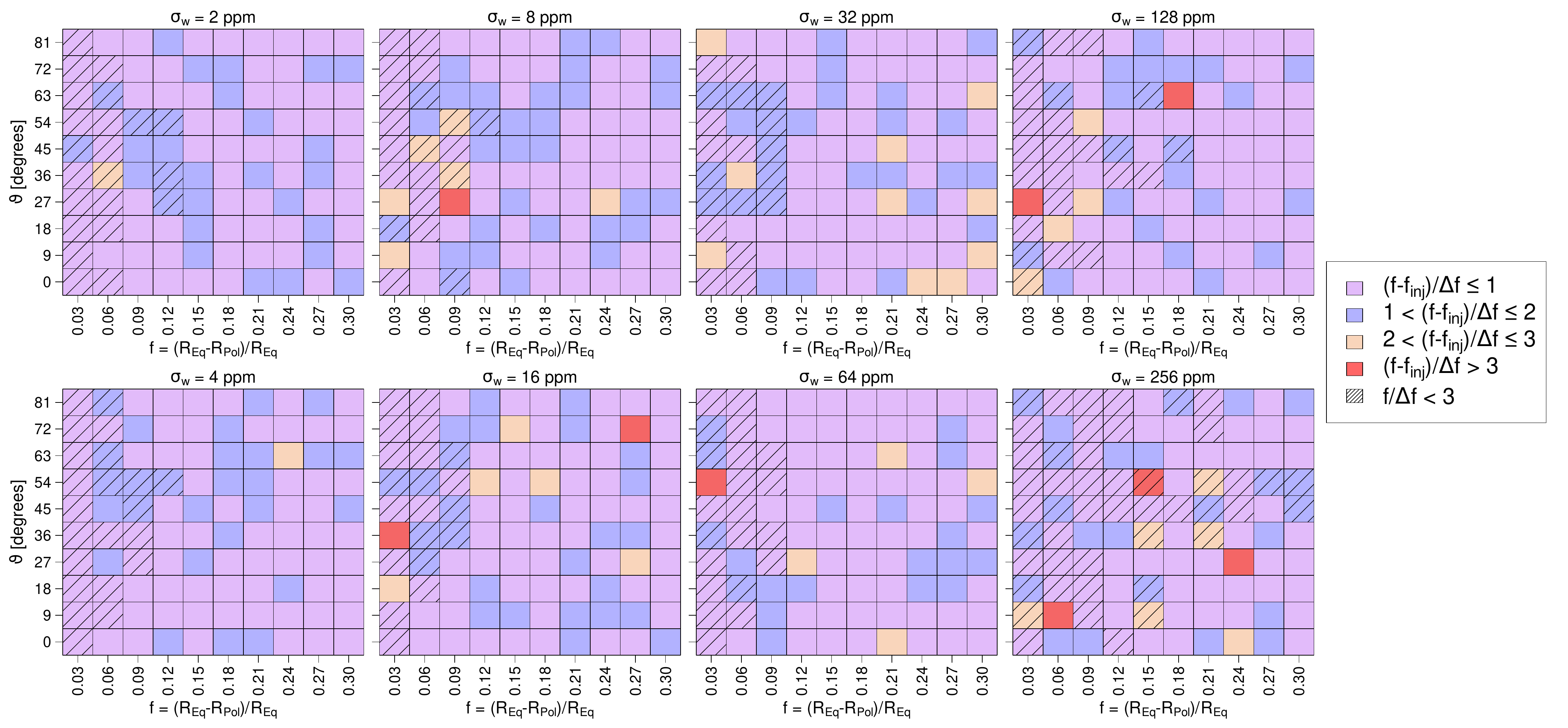}
    \caption{Precision of the retrieved oblateness parameter in the light curve which included red noise, for all eight $\sigma_w$ noise levels. Squares in the input $f$--$\vartheta$ grid are coloured based on the precision of the fitted $f$: purple if the retrieved parameter is within $1 \sigma$ of the injected value, blue if it is within $2 \sigma$, orange when it is within $3 \sigma$, and red otherwise. Shading is present when no significant detection of oblateness is made (i.e. when there is no $3 \sigma$ detection of $f$).}
    \label{fig:f_rel}
\end{figure}

\begin{figure}
    \centering
    \includegraphics[width=\linewidth]{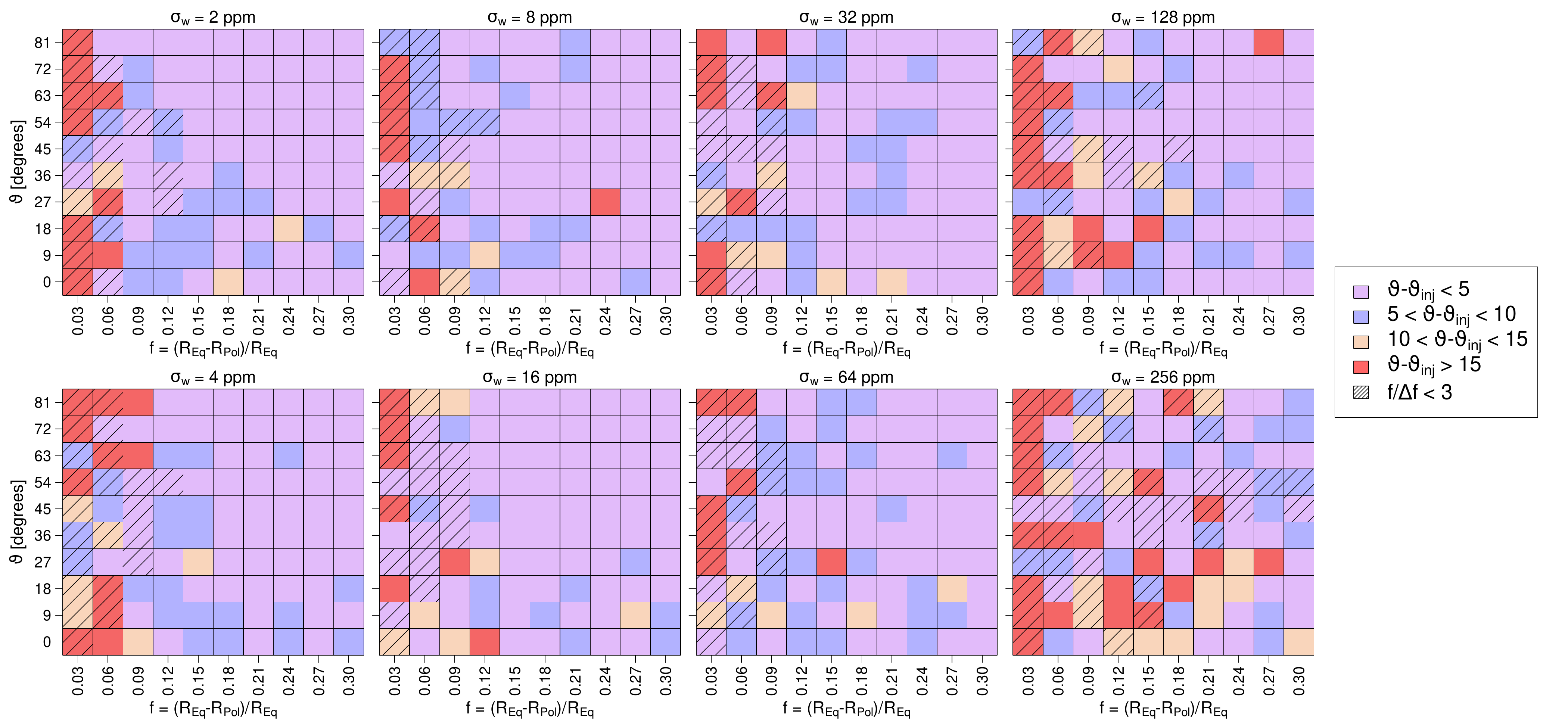}
    \caption{Accuracy of the retrieved obliquity parameter in the light curve which included red noise, for all eight $\sigma_w$ noise levels. Squares in the input $f$--$\vartheta$ grid are coloured based on the accuracy of the fitted $\vartheta$: purple if the retrieved parameter is within $5^\circ$ of the truth, blue if it is within $10^\circ$, orange when it is within $15^\circ$, and red otherwise. Shading is present when no significant detection of oblateness is made (i.e. when there is no $3 \sigma$ detection of $f$).}
    \label{fig:th_abs}
\end{figure}

\begin{figure}
    \centering
    \includegraphics[width=\linewidth]{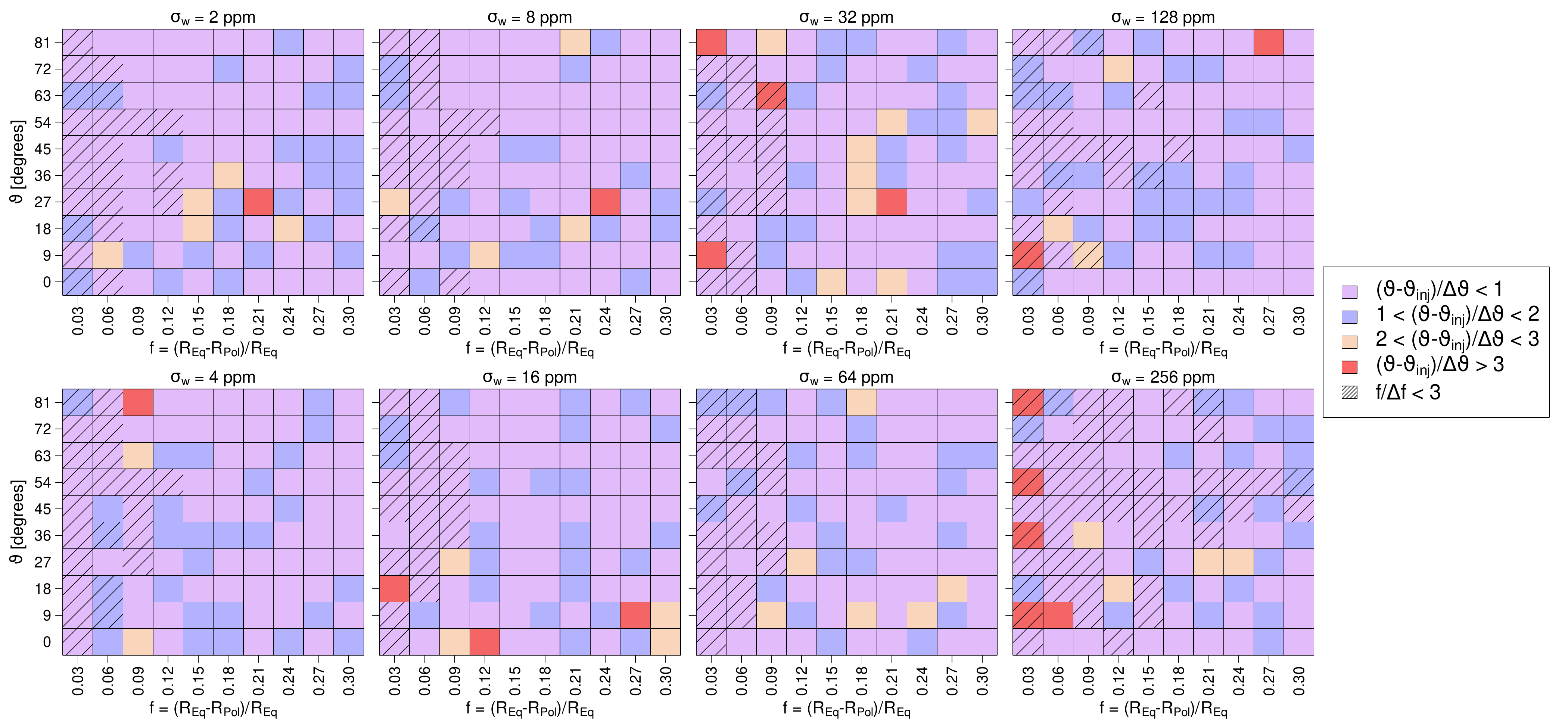}
    \caption{Precision of the retrieved obliquity parameter in the light curve which included red noise, for all eight $\sigma_w$ noise levels. Squares in the input $f$--$\vartheta$ grid are coloured based on the precision of the fitted $\theta$: purple if the retrieved parameter is within $1 \sigma$ of the injected value, blue if it is within $2 \sigma$, orange when it is within $3 \sigma$, and red otherwise. Shading is present when no significant detection of oblateness is made (i.e. when there is no $3 \sigma$ detection of $f$).}
    \label{fig:th_rel}
\end{figure}

\begin{table}[]
    \centering
        \caption{Parameters used in the analysis of the light curves of oblate planets. We applied uniform ($\mathcal{U}$) priors in every case, and in some cases, Gaussian priors ($\mathcal{N}$) were also in use.}
    \label{tab:parameter_setup}
    \begin{tabular}{c c c}
    \hline
    Parameter & \multicolumn{2}{c}{Prior} \\
    \hline
    \hline
       $P$ [days], fixed & \multicolumn{2}{l}{\textbf{$87$}} \\
        $\req$ & \multicolumn{2}{l}{$\mathcal{U}(0.0,1.0)$} \\
        $b$ & \multicolumn{2}{l}{$\mathcal{U}(-1.2,1.2)$} \\
        $a/R_\star$ & $\mathcal{U}(1.0, 199.0$) & $\mathcal{N}(73.26,0.04)$ \\
        $t_C$ [days] & \multicolumn{2}{l}{$\mathcal{U}(-0.01,0.01)$} \\
        $p_0$ [$100$ ppm]& \multicolumn{2}{l}{$\mathcal{U}(-100,100)$} \\
        $f$ & \multicolumn{2}{l}{$\mathcal{U}(0.0,0.4)$} \\
        $\vartheta$ [$^\circ$]& \multicolumn{2}{l}{$\mathcal{U}(-90.0,90.0)$} \\
        $\sigma_w$ [$100$ ppm]& \multicolumn{2}{l}{$\mathcal{U}(0.0,20000.0)$} \\
        $\sigma_r$ [$100$ ppm]& \multicolumn{2}{l}{$\mathcal{U}(0.0,10000.0)$} \\
        $u_A$ & $\mathcal{U}(-3.7,6.3)$ & $\mathcal{N}(1.30,0.01)$ \\
        $u_B$ & $\mathcal{U}(-3.7,6.3)$ & $\mathcal{N}(1.30,0.01)$ \\
       \hline
    \end{tabular}
\end{table}
We present the precision and accuracy of $f$ and $\theta$ on the $8 \times 10 \times 10$ grid on Figs. \ref{fig:f_abs} -- \ref{fig:th_rel}. We define a detection of oblateness (i.e. a transit light curve that cannot be explained well with a spherical planet) when the fitted oblateness parameter has a $3 \sigma$ significance ($f/\Delta f \geq 3$). Examining Figs. \ref{fig:f_abs} -- \ref{fig:th_rel}, we see that this criterion is more easily fulfilled for higher injected oblateness values. This is because the differences between the transit curve of a spherical planet and an oblate planet also increase with higher $f$ values, making the effect more easily detectable. In the cases with white noise levels lower than $256$~ppm, a $3 \sigma$ oblateness detection is not achieved in $141$ cases (out of $700$). At $\sigma_w = 256$~ppm, a $3 \sigma$ detection happened only in $54$ cases, suggesting that at this underlying white noise level, characterizing the oblateness of a transiting exoplanet is not feasible. The detection is also made easier when the rotation of the planet is either aligned with, or perpendicular to its orbital plane. This is because the overall signal introduced by the oblateness, as charaterized by the residual curve of an oblate planet and a spherical one (following Eq. \ref{eq:int_delta}), is larger at $\vartheta \approx 0^\circ$ or $ \vartheta \approx 90^\circ$ (Fig. \ref{fig:snr}).

The degeneracy between $b$ and $\vartheta$, as highlighted in Fig. \ref{fig:gl_transit}, is taken into account by taking $\left \vert \vartheta \right \vert$ if and only if $b < 0$. This way, we preserve the possible inaccurate retrievals of $\vartheta$ for a thorough assessment of the parameter retrieval.


The absolute differences between the injected ($f_{\rm inj}$) and retrieved oblateness ($f$) values are sorted into groups: $\left \vert f-f_{\rm inj} \right \vert \leq 0.02$, $0.02 < \left\vert f-f_{\rm inj} \right \vert  \leq0.04$, $0.04 < \left\vert f-f_{\rm inj} \right \vert  \leq0.06$, $\left\vert f-f_{\rm inj} \right \vert  > 0.06$. These groups are shown with color-code on Fig. \ref{fig:f_abs}. The arising pattern is complex. For noise levels below $256$~ppm, when a $3 \sigma$ detection is achieved, the retrieved oblateness is within $0.02$ of the truth in $346$ instances (out of $559$ cases, or $\approx 62\%$), within $0.04$ at $153$ grid positions ($\approx 27\%$), within $0.06$ of the input $41$ times ($\approx 7\%$), and $\left \vert f-f_{\rm inj} \right \vert > 0.06$ (the approximate oblateness of Jupiter) is only the case for $19$ ($3 \%$) of the light curves. In the $\sigma_w = 256$~ppm case, the recovered $f$ is only in a $0.02$ agreement with $f_{\rm inj}$ in $21$ instances ($0.39\%$) of the $3 \sigma$ detection cases, with the discrepancy being $> 0.06$ in $9$ light curves out of $54$ ($16\%$).

The relative differences between the fitted and input oblateness are shown on Fig. \ref{fig:f_rel}, color-coded according to their $z$ scores. When considering the $3 \sigma$ oblateness detections at all noise levels, we see a $1 \sigma$ agreement between $f_{\rm inj}$ and $f$ in 426 light curves ($\approx70\%$), a $2 \sigma$ agreement in $152$ cases ($\approx25\%$), $z = 3$ in $27$ instances ($\approx 4\%$), and a more than $3 \sigma$ discrepancy in only $8$ cases ($\approx 1\%$). The low number of $\geq 2 \sigma$ disagreements between the input and recovered values (even when a $3 \sigma$ detection is not achieved) suggests a robust uncertainty estimation of $f$. 



The absolute differences between the injected ($\vartheta_{\rm inj}$) and retrieved ($\vartheta$) obliquities are again divided into four categories: $\left \vert \vartheta-\vartheta_{\rm inj} \right \vert \leq 5^\circ$, $ 5^\circ < \left\vert \vartheta-\vartheta_{\rm inj} \right \vert  \leq 10^\circ$, $10^\circ < \left\vert \vartheta-\vartheta_{\rm inj} \right \vert  \leq 15^\circ$, $\left\vert \vartheta-\vartheta_{\rm inj} \right \vert  > 15^\circ$. The color-coded accuracy of $\vartheta$ over the entire grid of parameters in shown on Fig. \ref{fig:th_abs}. For noise levels below $256$~ppm, $413$ transits (out of the $559$ with $3\sigma$ oblateness detections) allow for the recovery of $\vartheta$ with $\leq 5^\circ$ accuracy, corresponding to $\approx 74\%$. 
We note that the divisions used to asses the accuracy of both the oblateness and obliquity parameters is arbitrary.

The precision of the retrieved $\vartheta$ parameters with respect to the injected $\vartheta_{\rm inj}$ parameters is divided into four groups based on the retrieved $z$-scores (Fig. \ref{fig:th_rel}). At every noise level, out of the $613$ cases where a $3 \sigma$ detection of oblateness was achieved, in $407$ analyses ($\approx 66\%$), the retrieved $\vartheta$ is in agreement with $\vartheta_{\rm inj}$ within the estimated uncertainties. In $160$ light curves ($\approx 26\%$), we find a $2 \sigma$ agreement, $3 \sigma$ agreement is found in $35$ cases ($\approx 6\%$), and a discrepancy $> 3 \sigma$ happens at $11$ grid positions ($\approx 2 \%$). Similarly to the case of the oblateness, the low number of $>2\sigma$ discrepancies suggest a robust uncertainty estimation for the oblateness.

We also show that the $3 \sigma$ detection of oblateness depends on the input oblateness itself. A significant detection of oblateness with $f_{\rm inj} = 0.03$ is only achieved in $8$ cases ($10\%$), however, these fitted $f$ values are in a $> 2 \sigma$ discrepancy with $f_{\rm inj}$. For Jupiter-like oblateness, we have a detection rate of $27.5\%$ ($22$ cases), and in $11$ of these, the retrieve oblateness agrees with $f_{\rm inj}$ within the estimated uncertainties, while the retrieved obliquities have a $\leq 1 \sigma$ agreement in $15$ out of these $22$ light curves. Saturn-like input oblateness values ($\approx 0.09$) can be detected at $47$ grid points ($\approx 59\%$). The fitted $f$ and $\vartheta$ values agree within $1 \sigma$ of their injected counterparts in $31$ and $30$ cases out of these $47$, respectively. At $f = 0.12$, a $3 \sigma$ oblateness detection is achieved at $68$ grid points ($85\%$) when including all noise levels. Excluding $\sigma_w = 256$~ppm, 
the detection rate increases to $90\%$.

\begin{figure}
    \centering
    \includegraphics[width = \linewidth]{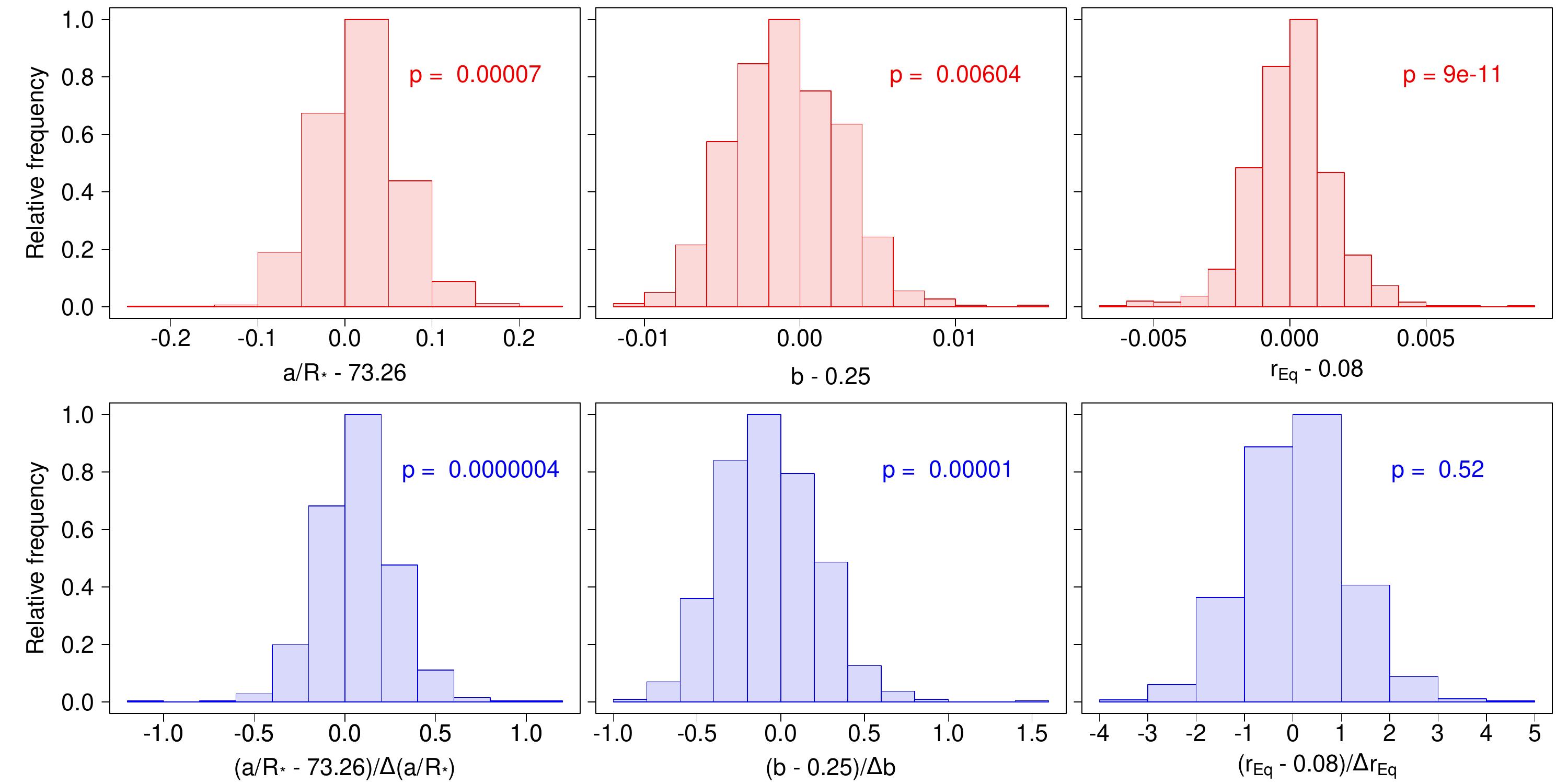}
    \caption{Distribution of the fitted semi-major axes (left), impact parameters (middle) and equatorial radii (right). The top row (red) shows the absolute deviations from the injected values, the bottom row (blue) shown the relative deviations from the injected values. We compare the distributions to Gaussian distributions via the Shapiro-Wilk test -- the resultant $p$ values are shown in every panel.}
    \label{fig:transitparams}
\end{figure}

\begin{figure}
    \centering
    \includegraphics[width=0.5\linewidth]{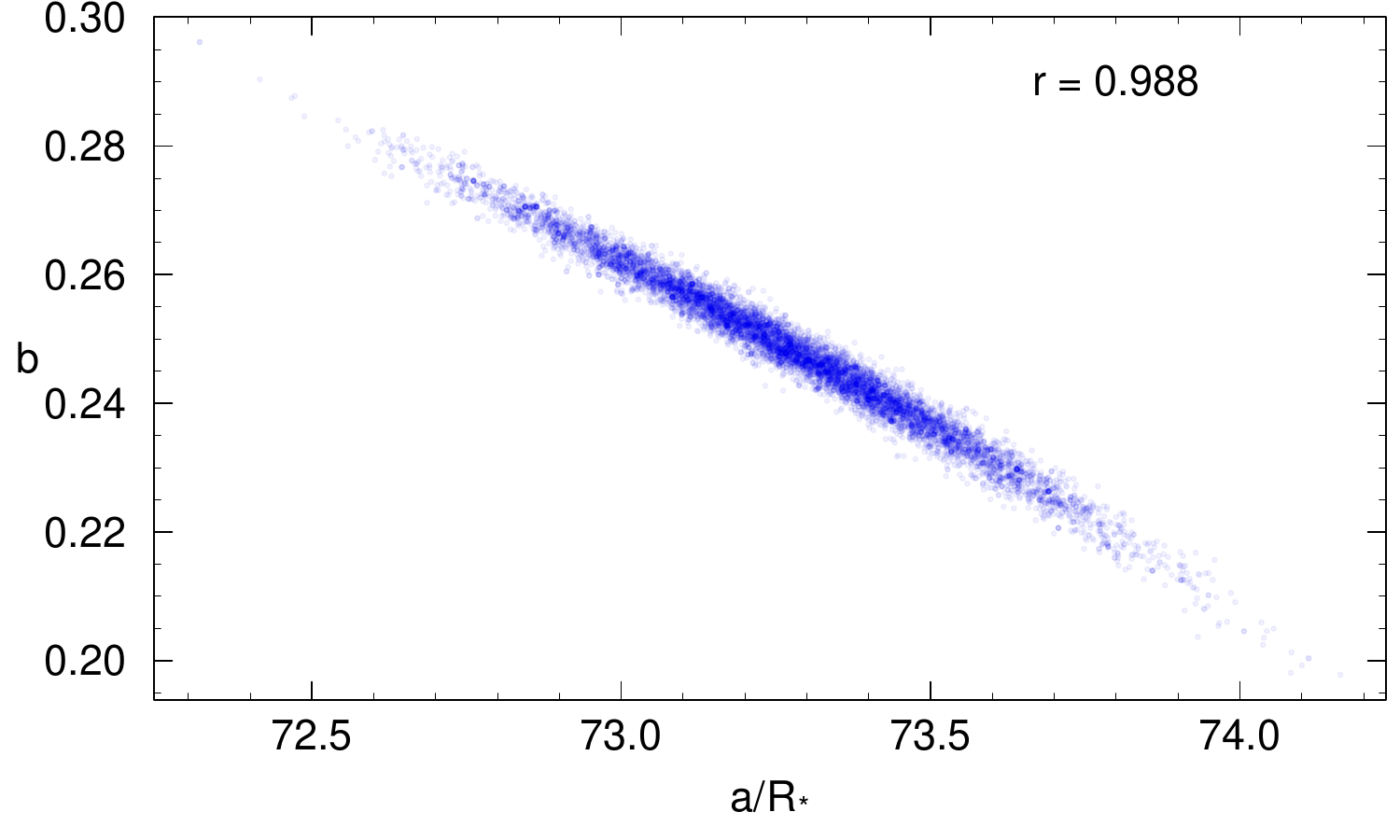}
    \caption{Posterior distribution in the $b$ -- $a/R_\star$ space from the $f = 0.06$, $\vartheta = 9^\circ$ case.}
    \label{fig:abdegen}
\end{figure}

We show the distribution of the fitted $a/R_\star$, $b$, and $\req$ values in Fig. \ref{fig:transitparams}, compiled for all $800$ light curves. The absolute deviations (Fig. \ref{fig:transitparams}, top row) are calculated by subtracting the input values ($73.26$, $0.25$, and $0.08$, for the three parameters, respectively) from the retrieved values. The relative deviations (Fig. \ref{fig:transitparams}) are then calculated as the ratio of the absolute deviations and the estimated uncertainties for every parameter, in every light curve. The median of the distribution of the semi-major axes is $0.012$, with a corresponding standard deviation of $0.049$. In total, $762$ retrieved $a/R_\star$ values are within $0.01$ of the injected semi-major axis. For the impact parameter, we find the median of the absolute deviations to be $-0.0009$, and a standard deviation of $0.0035$, after accounting for the $\vartheta$ -- $b$ degeneracy (Fig. \ref{fig:gl_transit}). We also find that $796$ retrieved values are between $0.24$ and $0.26$. For the equatorial radius, the median of the distribution is $0.0001$, accompanied by a standard deviation of $0.0015$. In $791$ cases, the fitted $\req$ is within $0.005$ of the input value of $0.08$. In the ideal case, we expect the standard deviation of the distribution of the relative discrepancies to be $1$ \citep[see][for more details]{2023A&A...675A.107K}. In the distributions seen in Fig. \ref{fig:transitparams}, the standard deviations of the relative distributions are $0.21$, $0.29$, and $1.07$ for $a/R_\star$, $b$, and $\req$, respectively. This implies that in the first two cases, the uncertainties are overestimated, likely as a consequence of the degeneracy between the two parameters and the oblateness. We show an example of the correlation between $a/R_\star$ and $b$ in Fig. \ref{fig:abdegen}. We find Pearson's $r = 0.988$ in that particular case, suggesting very high degeneracy between the two parameters. This is a well-known degeneracy in transit modeling \citep[see e.g.][]{2011A&A...531A..41C}.
For $\req$, the uncertainties are estimated correctly. We utilize the Shapiro-Wilk test for normality \citep{SHAPIRO1965} to compare all six distributions to Gaussian ones. If the resultant $p$-values are small, we can reject the null-hypothesis that the underlying parameters are drawn from a normal distribution. We expect that from a large-enough dataset, these distributions will behave asymptotically normally, if there are no autocorrelated effects in the residuals and if there are no degeneracies between the parameters. The wavelet-based noise filtering insures that the autocorrelated effects are negligible. If there are degeneracies between the parameters, we expect that the uncertainties are enlarged, so these distributions will be narrower than Gaussian distributions. Therefore, high $p$-values imply that there are degeneracies between the parameters. The analysis of the $p$-values (shown on Fig. \ref{fig:transitparams}) suggests that only the distribution of the relative $\req$ values are compatible with a Gaussian distribution.

\section{Discussion \& conclusion} \label{sec:conclusion}
We present a novel numerical approach for modelling the transit light curves of exoplanets whose shape is described by a biaxial ellipsoid due to their rapid rotation, following the recent surge of analytical methods \citep{2024arXiv240611644L, 2024arXiv241003449D, 2024JOSS....9.6972C}. We utilize the Gauss-Legendre quadrature, as described in Sect. \ref{sec:glq} for calculating the overlapping area between a the sky-projected contour (an ellipse) of the planet and a spherical star. This approach is then made available as a new module of TLCM \citep{2020MNRAS.496.4442C}.

In order to assess the efficiency and limitations of the model described in Sect. \ref{sect:transit_calc}, we compare it in the non-rotating (i.e. spherical) case to the well-established analytical approach of \cite{2002ApJ...580L.171M}. At larger $n$ number of integration points, the difference between the two methods is limited to the ingress and egress phases of the transit (Fig. \ref{fig:residuals}), where the residuals have a peak-to-peak amplitude on the order of several ppm. We introduce the ``area under the curve of the residuals'' quantity ($\Delta$) to obtain a more robust quantification of the difference between two light curves generated with different methods. In the specific cases tested in Sect. \ref{sec:tests}, measuring $\Delta$ provides a more in-depth understanding of the inherent noise that is the side effect of the numerical integration. We find that the error in accuracy is related to the volume of the planet ($\Delta \propto \req^3$, Fig. \ref{fig:rp_diff}), but the other basic transit parameters, the scaled semi-major axis, and the impact parameter (Figs. \ref{fig:a_diff}, \ref{fig:b_diff}), have more complex corresponding error terms. From the grid-based tests (shown in Table \ref{tab:transitparams}), we find that the minimal $\Delta$ of our numerical method can be achieved at $a/R_\star \approx 30$ and $b \approx 0.25$ (for the given $60$ second exposure time) suggesting one possible sweet spot for planning observations of planets with potentially detectable oblateness. 

We also show that the (as expected) the accuracy of the model increases with an increase in the $n$ number of integration points. This can be expressed both in the terms of the peak-to-peak amplitude of the residuals (Fig. \ref{fig:residuals}) and $\Delta$ (Fig. \ref{fig:effi}). Based on the assessment of the area under the curve of the residuals as it relates to the number of integration points, we find that $\Delta \propto n^{-2}$, as expected of a two-dimensional integration. In addition, we demonstrate that the runtime of our numerical modell is $\propto n^{0.1}$, and that it is $\approx 20\%$ faster than the baseline method of \cite{2002ApJ...580L.171M}
 for $n = 96$. 

We conduct a large-scale study to test the feasibility of retrieving the two parameters used in describing the transit of an oblate planet: the sky-projected oblateness ($f$) and the sky-projected obliquity ($\vartheta$). This is done by defining a grid of $f$ and $\vartheta$ values (yielding $100$ different configurations, Sect. \ref{sec:grid}), computing the transit light curves with $n = 96$, adding white noise where the standard deviation is taken from the $\sigma_w = \{2, 4, 8, 16, 32, 64, 128, 256\}$~ppm grid, and finally injecting time-correlated noise. Example light curves are shown in Fig. \ref{fig:demo_lcs}, used to demonstrate the noise levels. We then proceeded all $800$ light curves with TLCM, by setting $n = 96$ integration points.  The precision and accuracy of the retrieved oblateness and obliquity is shown on Figs. \ref{fig:f_abs}--\ref{fig:th_rel}. We utilize the prior knowledge of limb darkening, and place tight Gaussian priors on the two parameters describing in. Additionally, we also rely on the precise knowledge of the mean stellar density \citep{2017ApJ...835..173S} which may be obtainable via asteroseismology with \textit{PLATO} \citep{2014ExA....38..249R, 2024arXiv240605447R}, to constrain the scaled semi-major axis  (Eq. (\ref{eq:rho})). Even with these prior assumptions, we find that oblateness values below those of Jupiter ($f \approx 0.06$) is not retrievable at $3 \sigma$ with the method described here at any noise level. We also find that the oblateness and obliquity are detectable more easily for parallel or perpendicular projected spin-orbit angles.

We show that Saturn-like oblateness values ($f \approx 0.09$) are detectable in $\approx 59\%$ of the tested cases, and when a detection is made, the retrieved oblateness and obliquity is in a $1 \sigma$ agreement with the truth in $\approx 66\%$ and $\approx 64\%$ of the modelled light curves. We find that at higher oblateness ($f \geq 0.15$) and lower noise  ($\sigma_w < 64$~ppm) values, a significant detection of oblateness is guaranteed under the assumptions presented above. In these cases, the parameter retrieval is both precise and accurate. At $\sigma_w = 128$~ppm, and especially at $\sigma_w = 256$~ppm, even for the higher oblateness values, the fitted $f$ and $\vartheta$ values are less accurate, but are still precise, highlighting the robust uncertainty estimation in TLCM.

We find that the wavelet-based noise handling algorithm of \cite{2009ApJ...704...51C}, as implemented in TLCM \citep{2020MNRAS.496.4442C, 2023A&A...675A.106C}, is capable of accounting for the inherent noise of the analysis seen during the ingress and egress phases of a transit (Fig. \ref{fig:residuals}), thus negating one of the most prominent drawbacks of the light curve calculations described in Sect. \ref{sect:transit_calc}. 

The time-correlated noise was selected in an independent, random way for every light curve. Consequently, more problematic structures in it (such as the $\sigma_w = 8$~ppm case from Fig. \ref{fig:demo_lcs}) may degrade the precision and accuracy of the retrieved parameters. This is an unavoidable drawback of the test conducted here. For a more robust extrapolation of the parameter stability, a bootstrapping-like (or a Monte Carlo-like) process could be applied for each individual grid point, similarly to the work of \cite{2023A&A...675A.107K}.

The accuracy and precision plots in Figs. \ref{fig:f_abs}--\ref{fig:th_rel} can also serve as a lookup table for analyses of real observations. The noise levels can readily be scaled to the current $60$ second point-to-point scatter for various instruments and number of transits \citep[see e.g.][]{2019ApJ...878..119H}, however, one also has to ensure that there are sufficient observations during the ingress and egress of the planet (where the oblateness signal is the strongest) to be able to carry out the light curve modeling. If a $3 \sigma$ detection of oblateness is achieved, one can then tell by looking at Figs. \ref{fig:f_abs} -- \ref{fig:th_rel} how reliable the retrieved parameters and their uncertainties are, or how close they may be to the underlying phenomena. 

Further possible improvements can be achieved via the usage of the Love numbers of exoplanets. The fluid Love number $k_2$ can be measured via transit timing variations and radial velocity technique \citep{2019A&A...623A..45C,bernabo24,2025A&A...694A.233B}. The Love number $h_2$ has the biggest effect on the planetary shape and in case of hydrostatic equilibrium $h_2 = 1 + k_2$. The parameter $h_2$ specifies the shape deformation of the planet \citep{2019ApJ...878..119H} leaving only the obliquity as a free parameter.

Finally, we note that for a real detection of oblateness, a bright, quiet, Sun-like star is needed, for which the density can be measured with sufficiently high precision.


\begin{acknowledgements}
Project no. C1746651 has been implemented with the support provided by the Ministry of Culture and Innovation of Hungary from the National Research, Development and Innovation Fund, financed under the NVKDP-2021 funding scheme. This project has been supported by the SNN-147362 grant  of  the  Hungarian  Research,  Development  and  Innovation  Office  (NKFIH). The light curves files (both input and output), as well as any necessary configuration files, created for this project are available at \url{https://osf.io/496w2/?view_only=74c1306cd6044878859dc23c6e685b7a} for reproducibility.
\end{acknowledgements}

\software{TLCM \citep{2020MNRAS.496.4442C, 2023A&A...675A.106C, 2023A&A...675A.107K}}

\begin{contribution}
SzK, RSz, and GySzM defined the idea of the study. SzCs was responsible for creating the model. SzK, SzCs, and LMB carried out the tests. SzK and SzCs wrote and edited the manuscript. RSz also contributed to the interpretation.
\end{contribution}

\bibliographystyle{aasjournalv7}
\bibliography{refs}

\begin{thebibliography}{}
\expandafter\ifx\csname natexlab\endcsname\relax\def\natexlab#1{#1}\fi
\providecommand{\url}[1]{\href{#1}{#1}}
\providecommand{\dodoi}[1]{doi:~\href{http://doi.org/#1}{\nolinkurl{#1}}}
\providecommand{\doeprint}[1]{\href{http://ascl.net/#1}{\nolinkurl{http://ascl.net/#1}}}
\providecommand{\doarXiv}[1]{\href{https://arxiv.org/abs/#1}{\nolinkurl{https://arxiv.org/abs/#1}}}

\bibitem[{B. {Akinsanmi} {et~al.}(2019){Akinsanmi}, {Barros}, {Santos},
  {Correia}, {Maxted}, {Bou{\'e}}, \& {Laskar}}]{akinsamni19}
{Akinsanmi}, B., {Barros}, S.~C.~C., {Santos}, N.~C., {et~al.} 2019,
  \bibinfo{title}{{Detectability of shape deformation in short-period
  exoplanets},} \aap, 621, A117, \dodoi{10.1051/0004-6361/201834215}

\bibitem[{B. {Akinsanmi} {et~al.}(2024{\natexlab{a}}){Akinsanmi}, {Lendl},
  {Bou{\'e}}, \& {Barros}}]{2024A&A...682A..15A}
{Akinsanmi}, B., {Lendl}, M., {Bou{\'e}}, G., \& {Barros}, S. C.~C.
  2024{\natexlab{a}}, \bibinfo{title}{{Effects of tidal deformation on
  planetary phase curves},} \aap, 682, A15, \dodoi{10.1051/0004-6361/202347739}

\bibitem[{B. {Akinsanmi} {et~al.}(2024{\natexlab{b}}){Akinsanmi}, {Barros},
  {Lendl}, {Carone}, {Cubillos}, {Bekkelien}, {Fortier}, {Flor{\'e}n}, {Collier
  Cameron}, {Bou{\'e}}, {Bruno}, {Demory}, {Brandeker}, {Sousa}, {Wilson},
  {Deline}, {Bonfanti}, {Scandariato}, {Hooton}, {Correia}, {Demangeon},
  {Smith}, {Singh}, {Alibert}, {Alonso}, {Asquier}, {B{\'a}rczy}, {Barrado
  Navascues}, {Baumjohann}, {Beck}, {Beck}, {Benz}, {Billot}, {Bonfils},
  {Borsato}, {Broeg}, {Buder}, {Charnoz}, {Csizmadia}, {Davies}, {Deleuil},
  {Delrez}, {Ehrenreich}, {Erikson}, {Farinato}, {Fossati}, {Fridlund},
  {Gandolfi}, {Gillon}, {G{\"u}del}, {G{\"u}nther}, {Heitzmann}, {Helling},
  {Hoyer}, {Isaak}, {Kiss}, {Lam}, {Laskar}, {Lecavelier des Etangs}, {Magrin},
  {Maxted}, {Mecina}, {Mordasini}, {Nascimbeni}, {Olofsson}, {Ottensamer},
  {Pagano}, {Pall{\'e}}, {Peter}, {Piazza}, {Piotto}, {Pollacco}, {Queloz},
  {Ragazzoni}, {Rando}, {Rauer}, {Ribas}, {Santos}, {S{\'e}gransan}, {Simon},
  {Stalport}, {Szab{\'o}}, {Thomas}, {Udry}, {Van Grootel}, {Venturini},
  {Villaver}, \& {Walton}}]{2024A&A...685A..63A}
{Akinsanmi}, B., {Barros}, S.~C.~C., {Lendl}, M., {et~al.} 2024{\natexlab{b}},
  \bibinfo{title}{{The tidal deformation and atmosphere of WASP-12 b from its
  phase curve★},} \aap, 685, A63, \dodoi{10.1051/0004-6361/202348502}

\bibitem[{M. {Auvergne} {et~al.}(2009){Auvergne}, {Bodin}, {Boisnard}, {Buey},
  {Chaintreuil}, {Epstein}, {Jouret}, {Lam-Trong}, {Levacher}, {Magnan},
  {Perez}, {Plasson}, {Plesseria}, {Peter}, {Steller}, {Tiph{\`e}ne}, {Baglin},
  {Agogu{\'e}}, {Appourchaux}, {Barbet}, {Beaufort}, {Bellenger}, {Berlin},
  {Bernardi}, {Blouin}, {Boumier}, {Bonneau}, {Briet}, {Butler}, {Cautain},
  {Chiavassa}, {Costes}, {Cuvilho}, {Cunha-Parro}, {de Oliveira Fialho},
  {Decaudin}, {Defise}, {Djalal}, {Docclo}, {Drummond}, {Dupuis}, {Exil},
  {Faur{\'e}}, {Gaboriaud}, {Gamet}, {Gavalda}, {Grolleau}, {Gueguen},
  {Guivarc'h}, {Guterman}, {Hasiba}, {Huntzinger}, {Hustaix}, {Imbert},
  {Jeanville}, {Johlander}, {Jorda}, {Journoud}, {Karioty}, {Kerjean},
  {Lafond}, {Lapeyrere}, {Landiech}, {Larqu{\'e}}, {Laudet}, {Le Merrer},
  {Leporati}, {Leruyet}, {Levieuge}, {Llebaria}, {Martin}, {Mazy}, {Mesnager},
  {Michel}, {Moalic}, {Monjoin}, {Naudet}, {Neukirchner}, {Nguyen-Kim},
  {Ollivier}, {Orcesi}, {Ottacher}, {Oulali}, {Parisot}, {Perruchot},
  {Piacentino}, {Pinheiro da Silva}, {Platzer}, {Pontet}, {Pradines},
  {Quentin}, {Rohbeck}, {Rolland}, {Rollenhagen}, {Romagnan}, {Russ}, {Samadi},
  {Schmidt}, {Schwartz}, {Sebbag}, {Smit}, {Sunter}, {Tello}, {Toulouse},
  {Ulmer}, {Vandermarcq}, {Vergnault}, {Wallner}, {Waultier}, \&
  {Zanatta}}]{2009A&A...506..411A}
{Auvergne}, M., {Bodin}, P., {Boisnard}, L., {et~al.} 2009,
  \bibinfo{title}{{The CoRoT satellite in flight: description and
  performance},} \aap, 506, 411, \dodoi{10.1051/0004-6361/200810860}

\bibitem[{J.~W. {Barnes}(2009){Barnes}}]{2009ApJ...705..683B}
{Barnes}, J.~W. 2009, \bibinfo{title}{{Transit Lightcurves of Extrasolar
  Planets Orbiting Rapidly Rotating Stars},} \apj, 705, 683,
  \dodoi{10.1088/0004-637X/705/1/683}

\bibitem[{J.~W. {Barnes} \& J.~J. {Fortney}(2003){Barnes} \&
  {Fortney}}]{2003ApJ...588..545B}
{Barnes}, J.~W., \& {Fortney}, J.~J. 2003, \bibinfo{title}{{Measuring the
  Oblateness and Rotation of Transiting Extrasolar Giant Planets},} \apj, 588,
  545, \dodoi{10.1086/373893}

\bibitem[{W. {Benz} {et~al.}(2021){Benz}, {Broeg}, {Fortier}, {Rando}, {Beck},
  {Beck}, {Queloz}, {Ehrenreich}, {Maxted}, {Isaak}, {Billot}, {Alibert},
  {Alonso}, {Ant{\'o}nio}, {Asquier}, {Bandy}, {B{\'a}rczy}, {Barrado},
  {Barros}, {Baumjohann}, {Bekkelien}, {Bergomi}, {Biondi}, {Bonfils},
  {Borsato}, {Brandeker}, {Busch}, {Cabrera}, {Cessa}, {Charnoz}, {Chazelas},
  {Collier Cameron}, {Corral Van Damme}, {Cortes}, {Davies}, {Deleuil},
  {Deline}, {Delrez}, {Demangeon}, {Demory}, {Erikson}, {Farinato}, {Fossati},
  {Fridlund}, {Futyan}, {Gandolfi}, {Garcia Munoz}, {Gillon}, {Guterman},
  {Gutierrez}, {Hasiba}, {Heng}, {Hernandez}, {Hoyer}, {Kiss}, {Kovacs},
  {Kuntzer}, {Laskar}, {Lecavelier des Etangs}, {Lendl}, {L{\'o}pez}, {Lora},
  {Lovis}, {L{\"u}ftinger}, {Magrin}, {Malvasio}, {Marafatto}, {Michaelis}, {de
  Miguel}, {Modrego}, {Munari}, {Nascimbeni}, {Olofsson}, {Ottacher},
  {Ottensamer}, {Pagano}, {Palacios}, {Pall{\'e}}, {Peter}, {Piazza}, {Piotto},
  {Pizarro}, {Pollaco}, {Ragazzoni}, {Ratti}, {Rauer}, {Ribas}, {Rieder},
  {Rohlfs}, {Safa}, {Salatti}, {Santos}, {Scandariato}, {S{\'e}gransan},
  {Simon}, {Smith}, {Sordet}, {Sousa}, {Steller}, {Szab{\'o}}, {Szoke},
  {Thomas}, {Tschentscher}, {Udry}, {Van Grootel}, {Viotto}, {Walter},
  {Walton}, {Wildi}, \& {Wolter}}]{2021ExA....51..109B}
{Benz}, W., {Broeg}, C., {Fortier}, A., {et~al.} 2021, \bibinfo{title}{{The
  CHEOPS mission},} Experimental Astronomy, 51, 109,
  \dodoi{10.1007/s10686-020-09679-4}

\bibitem[{D. {Berardo} \& J. {de Wit}(2022){Berardo} \& {de
  Wit}}]{2022ApJ...935..178B}
{Berardo}, D., \& {de Wit}, J. 2022, \bibinfo{title}{{On the Effects of
  Planetary Oblateness on Exoplanet Studies},} \apj, 935, 178,
  \dodoi{10.3847/1538-4357/ac82b2}

\bibitem[{L.~M. {Bernab{\`o}} {et~al.}(2024){Bernab{\`o}}, {Csizmadia},
  {Smith}, {Rauer}, {Hatzes}, {Esposito}, {Gandolfi}, \& {Cabrera}}]{bernabo24}
{Bernab{\`o}}, L.~M., {Csizmadia}, S., {Smith}, A.~M.~S., {et~al.} 2024,
  \bibinfo{title}{{Evidence of apsidal motion and a possible co-moving
  companion star detected in the WASP-19 system},} \aap, 684, A78,
  \dodoi{10.1051/0004-6361/202346852}

\bibitem[{L.~M. {Bernab{\`o}} {et~al.}(2025){Bernab{\`o}}, {Csizmadia},
  {Smith}, {Harre}, {K{\'a}lm{\'a}n}, {Cabrera}, {Rauer}, {Gandolfi}, {Pino},
  {Ehrenreich}, \& {Hatzes}}]{2025A&A...694A.233B}
{Bernab{\`o}}, L.~M., {Csizmadia}, S., {Smith}, A.~M.~S., {et~al.} 2025,
  \bibinfo{title}{{Characterising WASP-43b's interior structure: Unveiling
  tidal decay and apsidal motion},} \aap, 694, A233,
  \dodoi{10.1051/0004-6361/202451994}

\bibitem[{U. {Bhowmick} \& V. {Khaire}(2024){Bhowmick} \&
  {Khaire}}]{2024AJ....168..243B}
{Bhowmick}, U., \& {Khaire}, V. 2024, \bibinfo{title}{{Yuti: A General-purpose
  Transit Simulator for Arbitrary Shaped Objects Orbiting Stars},} \aj, 168,
  243, \dodoi{10.3847/1538-3881/ad7d8d}

\bibitem[{W.~J. {Borucki} {et~al.}(2010){Borucki}, {Koch}, {Basri}, {Batalha},
  {Brown}, {Caldwell}, {Caldwell}, {Christensen-Dalsgaard}, {Cochran},
  {DeVore}, {Dunham}, {Dupree}, {Gautier}, {Geary}, {Gilliland}, {Gould},
  {Howell}, {Jenkins}, {Kondo}, {Latham}, {Marcy}, {Meibom}, {Kjeldsen},
  {Lissauer}, {Monet}, {Morrison}, {Sasselov}, {Tarter}, {Boss}, {Brownlee},
  {Owen}, {Buzasi}, {Charbonneau}, {Doyle}, {Fortney}, {Ford}, {Holman},
  {Seager}, {Steffen}, {Welsh}, {Rowe}, {Anderson}, {Buchhave}, {Ciardi},
  {Walkowicz}, {Sherry}, {Horch}, {Isaacson}, {Everett}, {Fischer}, {Torres},
  {Johnson}, {Endl}, {MacQueen}, {Bryson}, {Dotson}, {Haas}, {Kolodziejczak},
  {Van Cleve}, {Chandrasekaran}, {Twicken}, {Quintana}, {Clarke}, {Allen},
  {Li}, {Wu}, {Tenenbaum}, {Verner}, {Bruhweiler}, {Barnes}, \&
  {Prsa}}]{2010Sci...327..977B}
{Borucki}, W.~J., {Koch}, D., {Basri}, G., {et~al.} 2010,
  \bibinfo{title}{{Kepler Planet-Detection Mission: Introduction and First
  Results},} Science, 327, 977, \dodoi{10.1126/science.1185402}

\bibitem[{J.~A. {Carter} \& J.~N. {Winn}(2009){Carter} \&
  {Winn}}]{2009ApJ...704...51C}
{Carter}, J.~A., \& {Winn}, J.~N. 2009, \bibinfo{title}{{Parameter Estimation
  from Time-series Data with Correlated Errors: A Wavelet-based Method and its
  Application to Transit Light Curves},} \apj, 704, 51,
  \dodoi{10.1088/0004-637X/704/1/51}

\bibitem[{B. {Cassese} {et~al.}(2024){Cassese}, {Vega}, {Lu}, {Rice}, {Poddar},
  \& {Kipping}}]{2024JOSS....9.6972C}
{Cassese}, B., {Vega}, J., {Lu}, T., {et~al.} 2024,
  \bibinfo{title}{{squishyplanet: modeling transits of non-spherical exoplanets
  in JAX},} The Journal of Open Source Software, 9, 6972,
  \dodoi{10.21105/joss.06972}

\bibitem[{A. {Claret}(2004){Claret}}]{2004A&A...428.1001C}
{Claret}, A. 2004, \bibinfo{title}{{A new non-linear limb-darkening law for LTE
  stellar atmosphere models III. Sloan filters: Calculations for -5.0
  {\ensuremath{\leq}} log [M/H] {\ensuremath{\leq}} +1, 2000 K
  {\ensuremath{\leq}} T$_{eff}$ {\ensuremath{\leq}} 50 000 K at several surface
  gravities},} \aap, 428, 1001, \dodoi{10.1051/0004-6361:20041673}

\bibitem[{{\relax Sz}. {Csizmadia}(2020){Csizmadia}}]{2020MNRAS.496.4442C}
{Csizmadia}, {\relax Sz}. 2020, \bibinfo{title}{{The Transit and Light Curve
  Modeller},} \mnras, 496, 4442, \dodoi{10.1093/mnras/staa349}

\bibitem[{{\relax Sz}. {Csizmadia} {et~al.}(2019){Csizmadia}, {Hellard}, \&
  {Smith}}]{2019A&A...623A..45C}
{Csizmadia}, {\relax Sz}., {Hellard}, H., \& {Smith}, A.~M.~S. 2019,
  \bibinfo{title}{{An estimate of the k$_{2}$ Love number of WASP-18Ab from its
  radial velocity measurements},} \aap, 623, A45,
  \dodoi{10.1051/0004-6361/201834376}

\bibitem[{{\relax Sz}. {Csizmadia} {et~al.}(2023){Csizmadia}, {Smith},
  {K{\'a}lm{\'a}n}, {Cabrera}, {Klagyivik}, {Chaushev}, \&
  {Lam}}]{2023A&A...675A.106C}
{Csizmadia}, {\relax Sz}., {Smith}, A.~M.~S., {K{\'a}lm{\'a}n}, S., {et~al.}
  2023, \bibinfo{title}{{Power of wavelets in analyses of transit and phase
  curves in the presence of stellar variability and instrumental noise. I.
  Method and validation},} \aap, 675, A106, \dodoi{10.1051/0004-6361/202141302}

\bibitem[{{\relax Sz}. {Csizmadia} {et~al.}(2011){Csizmadia}, {Moutou},
  {Deleuil}, {Cabrera}, {Fridlund}, {Gandolfi}, {Aigrain}, {Alonso},
  {Almenara}, {Auvergne}, {Baglin}, {Barge}, {Bonomo}, {Bord{\'e}}, {Bouchy},
  {Bruntt}, {Carone}, {Carpano}, {Cavarroc}, {Cochran}, {Deeg}, {D{\'\i}az},
  {Dvorak}, {Endl}, {Erikson}, {Ferraz-Mello}, {Fruth}, {Gazzano}, {Gillon},
  {Guenther}, {Guillot}, {Hatzes}, {Havel}, {H{\'e}brard}, {Jehin}, {Jorda},
  {L{\'e}ger}, {Llebaria}, {Lammer}, {Lovis}, {MacQueen}, {Mazeh}, {Ollivier},
  {P{\"a}tzold}, {Queloz}, {Rauer}, {Rouan}, {Santerne}, {Schneider},
  {Tingley}, {Titz-Weider}, \& {Wuchterl}}]{2011A&A...531A..41C}
{Csizmadia}, {\relax Sz}., {Moutou}, C., {Deleuil}, M., {et~al.} 2011,
  \bibinfo{title}{{Transiting exoplanets from the CoRoT space mission. XVII.
  The hot Jupiter CoRoT-17b: a very old planet},} \aap, 531, A41,
  \dodoi{10.1051/0004-6361/201117009}

\bibitem[{R.~I. {Dawson} \& J.~A. {Johnson}(2012){Dawson} \&
  {Johnson}}]{2012ApJ...756..122D}
{Dawson}, R.~I., \& {Johnson}, J.~A. 2012, \bibinfo{title}{{The Photoeccentric
  Effect and Proto-hot Jupiters. I. Measuring Photometric Eccentricities of
  Individual Transiting Planets},} \apj, 756, 122,
  \dodoi{10.1088/0004-637X/756/2/122}

\bibitem[{M. {Deleuil} {et~al.}(2018){Deleuil}, {Aigrain}, {Moutou}, {Cabrera},
  {Bouchy}, {Deeg}, {Almenara}, {H{\'e}brard}, {Santerne}, {Alonso}, {Bonomo},
  {Bord{\'e}}, {Csizmadia}, {D{\`\i}az}, {Erikson}, {Fridlund}, {Gandolfi},
  {Guenther}, {Guillot}, {Guterman}, {Grziwa}, {Hatzes}, {L{\'e}ger}, {Mazeh},
  {Ofir}, {Ollivier}, {P{\"a}tzold}, {Parviainen}, {Rauer}, {Rouan},
  {Schneider}, {Titz-Weider}, {Tingley}, \& {Weingrill}}]{deleuil18}
{Deleuil}, M., {Aigrain}, S., {Moutou}, C., {et~al.} 2018,
  \bibinfo{title}{{Planets, candidates, and binaries from the CoRoT/Exoplanet
  programme. The CoRoT transit catalogue},} \aap, 619, A97,
  \dodoi{10.1051/0004-6361/201731068}

\bibitem[{S. {Dholakia} {et~al.}(2024){Dholakia}, {Dholakia}, \&
  {Pope}}]{2024arXiv241003449D}
{Dholakia}, S., {Dholakia}, S., \& {Pope}, B. J.~S. 2024, \bibinfo{title}{{A
  General, Differentiable Transit Model for Ellipsoidal Occulters: Derivation,
  Application, and Forecast of Planetary Oblateness and Obliquity Constraints
  with JWST},} arXiv e-prints, arXiv:2410.03449,
  \dodoi{10.48550/arXiv.2410.03449}

\bibitem[{J. {Diaz-Cordoves} \& A. {Gimenez}(1992){Diaz-Cordoves} \&
  {Gimenez}}]{1992A&A...259..227D}
{Diaz-Cordoves}, J., \& {Gimenez}, A. 1992, \bibinfo{title}{{A new nonlinear
  approximation to the limb-darkening of hot stars},} \aap, 259, 227

\bibitem[{E.~B. {Ford}(2006){Ford}}]{2006ApJ...642..505F}
{Ford}, E.~B. 2006, \bibinfo{title}{{Improving the Efficiency of Markov Chain
  Monte Carlo for Analyzing the Orbits of Extrasolar Planets},} \apj, 642, 505,
  \dodoi{10.1086/500802}

\bibitem[{J.~P. {Gardner} {et~al.}(2006){Gardner}, {Mather}, {Clampin},
  {Doyon}, {Greenhouse}, {Hammel}, {Hutchings}, {Jakobsen}, {Lilly}, {Long},
  {Lunine}, {McCaughrean}, {Mountain}, {Nella}, {Rieke}, {Rieke}, {Rix},
  {Smith}, {Sonneborn}, {Stiavelli}, {Stockman}, {Windhorst}, \&
  {Wright}}]{2006SSRv..123..485G}
{Gardner}, J.~P., {Mather}, J.~C., {Clampin}, M., {et~al.} 2006,
  \bibinfo{title}{{The James Webb Space Telescope},} \ssr, 123, 485,
  \dodoi{10.1007/s11214-006-8315-7}

\bibitem[{H. {Hellard} {et~al.}(2019){Hellard}, {Csizmadia}, {Padovan},
  {Rauer}, {Cabrera}, {Sohl}, {Spohn}, \& {Breuer}}]{2019ApJ...878..119H}
{Hellard}, H., {Csizmadia}, S., {Padovan}, S., {et~al.} 2019,
  \bibinfo{title}{{Retrieval of the Fluid Love Number k $_{2}$ in Exoplanetary
  Transit Curves},} \apj, 878, 119, \dodoi{10.3847/1538-4357/ab2048}

\bibitem[{H. {Hellard} {et~al.}(2020){Hellard}, {Csizmadia}, {Padovan}, {Sohl},
  \& {Rauer}}]{2020ApJ...889...66H}
{Hellard}, H., {Csizmadia}, S., {Padovan}, S., {Sohl}, F., \& {Rauer}, H. 2020,
  \bibinfo{title}{{HST/STIS Capability for Love Number Measurement of
  WASP-121b},} \apj, 889, 66, \dodoi{10.3847/1538-4357/ab616e}

\bibitem[{D. {Hestroffer}(1997){Hestroffer}}]{1997A&A...327..199H}
{Hestroffer}, D. 1997, \bibinfo{title}{{Centre to limb darkening of stars. New
  model and application to stellar interferometry.},} \aap, 327, 199

\bibitem[{{\relax Sz}. {K{\'a}lm{\'a}n} {et~al.}(2024){K{\'a}lm{\'a}n},
  {Csizmadia}, {Simon}, {Lam}, {Deline}, {Harre}, \&
  {Szab{\'o}}}]{kalman24_agent}
{K{\'a}lm{\'a}n}, {\relax Sz}., {Csizmadia}, S., {Simon}, A.~E., {et~al.} 2024,
  \bibinfo{title}{{Modelling the light curves of transiting exomoons: a
  zero-order photodynamic agent added to the Transit and Light Curve
  Modeller},} \mnras, 528, L66, \dodoi{10.1093/mnrasl/slad169}

\bibitem[{{\relax Sz}. {K{\'a}lm{\'a}n} {et~al.}(2023){K{\'a}lm{\'a}n},
  {Szab{\'o}}, \& {Csizmadia}}]{2023A&A...675A.107K}
{K{\'a}lm{\'a}n}, {\relax Sz}., {Szab{\'o}}, G.~M., \& {Csizmadia}, S. 2023,
  \bibinfo{title}{{Power of wavelets in analyses of transit and phase curves in
  the presence of stellar variability and instrumental noise. II. Accuracy of
  the transit parameters},} \aap, 675, A107,
  \dodoi{10.1051/0004-6361/202143017}

\bibitem[{D.~A. {Klinglesmith} \& S. {Sobieski}(1970){Klinglesmith} \&
  {Sobieski}}]{1970AJ.....75..175K}
{Klinglesmith}, D.~A., \& {Sobieski}, S. 1970, \bibinfo{title}{{Nonlinear Limb
  Darkening for Early-Type Stars},} \aj, 75, 175, \dodoi{10.1086/110960}

\bibitem[{Q. {Liu} {et~al.}(2024){Liu}, {Zhu}, {Zhou}, {Hu}, {Lin}, {Dai},
  {Masuda}, \& {Wang}}]{2024arXiv240611644L}
{Liu}, Q., {Zhu}, W., {Zhou}, Y., {et~al.} 2024, \bibinfo{title}{{Detecting
  Planetary Oblateness in the Era of JWST: A Case Study of Kepler-167e},} arXiv
  e-prints, arXiv:2406.11644, \dodoi{10.48550/arXiv.2406.11644}

\bibitem[{K. {Mandel} \& E. {Agol}(2002){Mandel} \&
  {Agol}}]{2002ApJ...580L.171M}
{Mandel}, K., \& {Agol}, E. 2002, \bibinfo{title}{{Analytic Light Curves for
  Planetary Transit Searches},} \apjl, 580, L171, \dodoi{10.1086/345520}

\bibitem[{F. {Matuszewski} {et~al.}(2023){Matuszewski}, {Nettelmann},
  {Cabrera}, {B{\"o}rner}, \& {Rauer}}]{2023A&A...677A.133M}
{Matuszewski}, F., {Nettelmann}, N., {Cabrera}, J., {B{\"o}rner}, A., \&
  {Rauer}, H. 2023, \bibinfo{title}{{Estimating the number of planets that
  PLATO can detect},} \aap, 677, A133, \dodoi{10.1051/0004-6361/202245287}

\bibitem[{M.~W. {McElwain} {et~al.}(2023){McElwain}, {Feinberg}, {Perrin},
  {Clampin}, {Mountain}, {Lallo}, {Lajoie}, {Kimble}, {Bowers}, {Stark},
  {Acton}, {Atkinson}, {Barinek}, {Barto}, {Basinger}, {Beck}, {Bergkoetter},
  {Bluth}, {Boucarut}, {Brady}, {Brooks}, {Brown}, {Byard}, {Carey},
  {Carrasquilla}, {Chae}, {Chaney}, {Chayer}, {Chonis}, {Cohen}, {Cole},
  {Comeau}, {Coon}, {Coppock}, {Coyle}, {Dean}, {Dziak}, {Eisenhower},
  {Flagey}, {Franck}, {Gallagher}, {Gilman}, {Glassman}, {Green}, {Grieco},
  {Haase}, {Hadjimichael}, {Hagopian}, {Hahn}, {Hartig}, {Havey}, {Hayden},
  {Hellekson}, {Hicks}, {Holfeltz}, {Howard}, {Huguet}, {Jahne}, {Johnson},
  {Johnston}, {Jurling}, {Kegley}, {Kennard}, {Keski-Kuha}, {Knight}, {Kulp},
  {Levi}, {Levine}, {Lightsey}, {Luetgens}, {Mather}, {Matthews}, {McKay},
  {Mehalick}, {Mel{\'e}ndez}, {Mosier}, {Murphy}, {Nelan}, {Niedner}, {Nol},
  {Ohara}, {Ohl}, {Olczak}, {Osborne}, {Park}, {Perrygo}, {Pueyo}, {Redding},
  {Regan}, {Reynolds}, {Rifelli}, {Rigby}, {Sabatke}, {Saif}, {Scorse}, {Seo},
  {Shi}, {Sigrist}, {Smith}, {Smith}, {Smith}, {Sohn}, {Stahl}, {Telfer},
  {Terlecki}, {Texter}, {Van Buren}, {Van Campen}, {Vila}, {Voyton}, {Waldman},
  {Walker}, {Weiser}, {Wells}, {West}, {Whitman}, {Wolf}, \&
  {Zielinski}}]{2023PASP..135e8001M}
{McElwain}, M.~W., {Feinberg}, L.~D., {Perrin}, M.~D., {et~al.} 2023,
  \bibinfo{title}{{The James Webb Space Telescope Mission: Optical Telescope
  Element Design, Development, and Performance},} \pasp, 135, 058001,
  \dodoi{10.1088/1538-3873/acada0}

\bibitem[{E.~A. {Milne}(1921){Milne}}]{1921MNRAS..81..361M}
{Milne}, E.~A. 1921, \bibinfo{title}{{Radiative equilibrium in the outer layers
  of a star},} \mnras, 81, 361, \dodoi{10.1093/mnras/81.5.361}

\bibitem[{E.~M. {Price} {et~al.}(2025){Price}, {Becker}, {de Beurs}, {Rogers},
  \& {Vanderburg}}]{2025ApJ...981L...7P}
{Price}, E.~M., {Becker}, J., {de Beurs}, Z.~L., {Rogers}, L.~A., \&
  {Vanderburg}, A. 2025, \bibinfo{title}{{A Long Spin Period for a
  Sub-Neptune-mass Exoplanet},} \apjl, 981, L7,
  \dodoi{10.3847/2041-8213/adb42b}

\bibitem[{H. {Rauer} {et~al.}(2014){Rauer}, {Catala}, {Aerts}, {Appourchaux},
  {Benz}, {Brandeker}, {Christensen-Dalsgaard}, {Deleuil}, {Gizon}, {Goupil},
  {G{\"u}del}, {Janot-Pacheco}, {Mas-Hesse}, {Pagano}, {Piotto}, {Pollacco},
  {Santos}, {Smith}, {Su{\'a}rez}, {Szab{\'o}}, {Udry}, {Adibekyan}, {Alibert},
  {Almenara}, {Amaro-Seoane}, {Eiff}, {Asplund}, {Antonello}, {Barnes},
  {Baudin}, {Belkacem}, {Bergemann}, {Bihain}, {Birch}, {Bonfils}, {Boisse},
  {Bonomo}, {Borsa}, {Brand{\~a}o}, {Brocato}, {Brun}, {Burleigh}, {Burston},
  {Cabrera}, {Cassisi}, {Chaplin}, {Charpinet}, {Chiappini}, {Church},
  {Csizmadia}, {Cunha}, {Damasso}, {Davies}, {Deeg}, {D{\'\i}az}, {Dreizler},
  {Dreyer}, {Eggenberger}, {Ehrenreich}, {Eigm{\"u}ller}, {Erikson}, {Farmer},
  {Feltzing}, {de Oliveira Fialho}, {Figueira}, {Forveille}, {Fridlund},
  {Garc{\'\i}a}, {Giommi}, {Giuffrida}, {Godolt}, {Gomes da Silva}, {Granzer},
  {Grenfell}, {Grotsch-Noels}, {G{\"u}nther}, {Haswell}, {Hatzes},
  {H{\'e}brard}, {Hekker}, {Helled}, {Heng}, {Jenkins}, {Johansen},
  {Khodachenko}, {Kislyakova}, {Kley}, {Kolb}, {Krivova}, {Kupka}, {Lammer},
  {Lanza}, {Lebreton}, {Magrin}, {Marcos-Arenal}, {Marrese}, {Marques},
  {Martins}, {Mathis}, {Mathur}, {Messina}, {Miglio}, {Montalban}, {Montalto},
  {Monteiro}, {Moradi}, {Moravveji}, {Mordasini}, {Morel}, {Mortier},
  {Nascimbeni}, {Nelson}, {Nielsen}, {Noack}, {Norton}, {Ofir}, {Oshagh},
  {Ouazzani}, {P{\'a}pics}, {Parro}, {Petit}, {Plez}, {Poretti}, {Quirrenbach},
  {Ragazzoni}, {Raimondo}, {Rainer}, {Reese}, {Redmer}, {Reffert},
  {Rojas-Ayala}, {Roxburgh}, {Salmon}, {Santerne}, {Schneider}, {Schou},
  {Schuh}, {Schunker}, {Silva-Valio}, {Silvotti}, {Skillen}, {Snellen}, {Sohl},
  {Sousa}, {Sozzetti}, {Stello}, {Strassmeier}, {{\v{S}}vanda}, {Szab{\'o}},
  {Tkachenko}, {Valencia}, {Van Grootel}, {Vauclair}, {Ventura}, {Wagner},
  {Walton}, {Weingrill}, {Werner}, {Wheatley}, \&
  {Zwintz}}]{2014ExA....38..249R}
{Rauer}, H., {Catala}, C., {Aerts}, C., {et~al.} 2014, \bibinfo{title}{{The
  PLATO 2.0 mission},} Experimental Astronomy, 38, 249,
  \dodoi{10.1007/s10686-014-9383-4}

\bibitem[{H. {Rauer} {et~al.}(2024){Rauer}, {Aerts}, {Cabrera}, {Deleuil},
  {Erikson}, {Gizon}, {Goupil}, {Heras}, {Lorenzo-Alvarez}, {Marliani},
  {Martin-Garcia}, {Mas-Hesse}, {O'Rourke}, {Osborn}, {Pagano}, {Piotto},
  {Pollacco}, {Ragazzoni}, {Ramsay}, {Udry}, {Appourchaux}, {Benz},
  {Brandeker}, {G{\"u}del}, {Janot-Pacheco}, {Kabath}, {Kjeldsen}, {Min},
  {Santos}, {Smith}, {Suarez}, {Werner}, {Aboudan}, {Abreu}, {Acu a}, {Adams},
  {Adibekyan}, {Affer}, {Agneray}, {Agnor}, {Aguirre B{\o}rsen-Koch}, {Ahmed},
  {Aigrain}, {Al-Bahlawan}, {Alcacera Gil}, {Alei}, {Alencar}, {Alexander},
  {Alfonso-Garz{\'o}n}, {Alibert}, {Allende Prieto}, {Almeida}, {Alonso
  Sobrino}, {Altavilla}, {Althaus}, {Alonso Alvarez Trujillo}, {Amarsi},
  {Ammler-von Eiff}, {Am{\^o}res}, {Andrade}, {Antoniadis-Karnavas},
  {Ant{\'o}nio}, {Aparicio del Moral}, {Appolloni}, {Arena}, {Armstrong},
  {Aroca Aliaga}, {Asplund}, {Audenaert}, {Auricchio}, {Avelino}, {Baeke},
  {Bailli{\'e}}, {Balado}, {Ballber Balaguer{\'o}}, {Balestra}, {Ball},
  {Ballans}, {Ballot}, {Barban}, {Barbary}, {Barbieri}, {Barcel{\'o} Forteza},
  {Barker}, {Barklem}, {Barnes}, {Barrado Navascues}, {Barragan}, {Baruteau},
  {Basu}, {Baudin}, {Baumeister}, {Bayliss}, {Bazot}, {Beck}, {Bedding},
  {Belkacem}, {Bellinger}, {Benatti}, {Benomar}, {B{\'e}rard}, {Bergemann},
  {Bergomi}, {Bernardo}, {Biazzo}, {Bignamini}, {Bigot}, {Billot}, {Binet},
  {Biondi}, {Biondi}, {Birch}, {Bitsch}, {Bluhm Ceballos}, {B{\'o}di},
  {Bogn{\'a}r}, {Boisse}, {Bolmont}, {Bonanno}, {Bonavita}, {Bonfanti},
  {Bonfils}, {Bonito}, {Bonomo}, {B{\"o}rner}, {Boro Saikia}, {Borreguero
  Mart{\'\i}n}, {Borsa}, {Borsato}, {Bossini}, {Bouchy}, {Bou{\'e}},
  {Boufleur}, {Boumier}, {Bourrier}, {Bowman}, {Bozzo}, {Bradley}, {Bray},
  {Bressan}, {Breton}, {Brienza}, {Brito}, {Brogi}, {Brown}, {Brown}, {Brun},
  {Bruno}, {Bruns}, {Buchhave}, {Bugnet}, {Buldgen}, {Burgess}, {Busatta},
  {Busso}, {Buzasi}, {Caballero}, {Cabral}, {Cabrero Gomez}, {Calderone},
  {Cameron}, {Cameron}, {Campante}, {Campos Gestal}, {Canto Martins}, {Cara},
  {Carone}, {Carrasco}, {Casagrande}, {Casewell}, {Cassisi}, {Castellani},
  {Castro}, {Catala}, {Catal{\'a}n Fern{\'a}ndez}, {Catelan}, {Cegla},
  {Cerruti}, {Cessa}, {Chadid}, {Chaplin}, {Charpinet}, {Chiappini},
  {Chiarucci}, {Chiavassa}, {Chinellato}, {Chirulli}, {Christensen-Dalsgaard},
  {Church}, {Claret}, {Clarke}, {Claudi}, {Clermont}, {Coelho}, {Coelho},
  {Cogato}, {Colom{\'e}}, {Condamin}, {Conde Garc{\'\i}a}, \&
  {Conseil}}]{2024arXiv240605447R}
{Rauer}, H., {Aerts}, C., {Cabrera}, J., {et~al.} 2024, \bibinfo{title}{{The
  PLATO Mission},} arXiv e-prints, arXiv:2406.05447,
  \dodoi{10.48550/arXiv.2406.05447}

\bibitem[{E. {Rein} \& A. {Ofir}(2019){Rein} \& {Ofir}}]{2019MNRAS.490.1111R}
{Rein}, E., \& {Ofir}, A. 2019, \bibinfo{title}{{Fast and precise light-curve
  model for transiting exoplanets with rings},} \mnras, 490, 1111,
  \dodoi{10.1093/mnras/stz2556}

\bibitem[{G.~R. {Ricker} {et~al.}(2015){Ricker}, {Winn}, {Vanderspek},
  {Latham}, {Bakos}, {Bean}, {Berta-Thompson}, {Brown}, {Buchhave}, {Butler},
  {Butler}, {Chaplin}, {Charbonneau}, {Christensen-Dalsgaard}, {Clampin},
  {Deming}, {Doty}, {De Lee}, {Dressing}, {Dunham}, {Endl}, {Fressin}, {Ge},
  {Henning}, {Holman}, {Howard}, {Ida}, {Jenkins}, {Jernigan}, {Johnson},
  {Kaltenegger}, {Kawai}, {Kjeldsen}, {Laughlin}, {Levine}, {Lin}, {Lissauer},
  {MacQueen}, {Marcy}, {McCullough}, {Morton}, {Narita}, {Paegert}, {Palle},
  {Pepe}, {Pepper}, {Quirrenbach}, {Rinehart}, {Sasselov}, {Sato}, {Seager},
  {Sozzetti}, {Stassun}, {Sullivan}, {Szentgyorgyi}, {Torres}, {Udry}, \&
  {Villasenor}}]{2015JATIS...1a4003R}
{Ricker}, G.~R., {Winn}, J.~N., {Vanderspek}, R., {et~al.} 2015,
  \bibinfo{title}{{Transiting Exoplanet Survey Satellite (TESS)},} Journal of
  Astronomical Telescopes, Instruments, and Systems, 1, 014003,
  \dodoi{10.1117/1.JATIS.1.1.014003}

\bibitem[{H.~N. {Russell}(1912){Russell}}]{russell12}
{Russell}, H.~N. 1912, \bibinfo{title}{{On the Determination of the Orbital
  Elements of Eclipsing Variable Stars. II.},} \apj, 36, 54,
  \dodoi{10.1086/141952}

\bibitem[{S. {Seager} \& G. {Mall{\'e}n-Ornelas}(2003){Seager} \&
  {Mall{\'e}n-Ornelas}}]{2003ApJ...585.1038S}
{Seager}, S., \& {Mall{\'e}n-Ornelas}, G. 2003, \bibinfo{title}{{A Unique
  Solution of Planet and Star Parameters from an Extrasolar Planet Transit
  Light Curve},} \apj, 585, 1038, \dodoi{10.1086/346105}

\bibitem[{S.~S. Shapiro \& M.~B. Wilk(1965)Shapiro \& Wilk}]{SHAPIRO1965}
Shapiro, S.~S., \& Wilk, M.~B. 1965, \bibinfo{title}{An analysis of variance
  test for normality (complete samples),} Biometrika, 52, 591,
  \dodoi{10.1093/biomet/52.3-4.591}

\bibitem[{V. {Silva Aguirre} {et~al.}(2017){Silva Aguirre}, {Lund}, {Antia},
  {Ball}, {Basu}, {Christensen-Dalsgaard}, {Lebreton}, {Reese}, {Verma},
  {Casagrande}, {Justesen}, {Mosumgaard}, {Chaplin}, {Bedding}, {Davies},
  {Handberg}, {Houdek}, {Huber}, {Kjeldsen}, {Latham}, {White}, {Coelho},
  {Miglio}, \& {Rendle}}]{2017ApJ...835..173S}
{Silva Aguirre}, V., {Lund}, M.~N., {Antia}, H.~M., {et~al.} 2017,
  \bibinfo{title}{{Standing on the Shoulders of Dwarfs: the Kepler
  Asteroseismic LEGACY Sample. II.Radii, Masses, and Ages},} \apj, 835, 173,
  \dodoi{10.3847/1538-4357/835/2/173}

\bibitem[{D.~K. {Sing}(2010){Sing}}]{2010A&A...510A..21S}
{Sing}, D.~K. 2010, \bibinfo{title}{{Stellar limb-darkening coefficients for
  CoRot and Kepler},} \aap, 510, A21, \dodoi{10.1051/0004-6361/200913675}

\bibitem[{{\relax Gy}.~M. {Szab{\'o}} {et~al.}(2022){Szab{\'o}},
  {K{\'a}lm{\'a}n}, {Pribulla}, {Claret}, {Mugnai}, {Pascale}, {Waltham},
  {Borsato}, {Garai}, \& {Szab{\'o}}}]{2022ExA....53..607S}
{Szab{\'o}}, {\relax Gy}.~M., {K{\'a}lm{\'a}n}, S., {Pribulla}, T., {et~al.}
  2022, \bibinfo{title}{{High-precision photometry with Ariel},} Experimental
  Astronomy, 53, 607, \dodoi{10.1007/s10686-021-09777-x}

\bibitem[{G. {Tinetti} {et~al.}(2022){Tinetti}, {Eccleston}, {Lueftinger},
  {Salvignol}, {Fahmy}, \& {Alves de Oliveira}}]{2022EPSC...16.1114T}
{Tinetti}, G., {Eccleston}, P., {Lueftinger}, T., {et~al.} 2022, in European
  Planetary Science Congress, EPSC2022--1114, \dodoi{10.5194/epsc2022-1114}

\bibitem[{G. {Tinetti} {et~al.}(2018){Tinetti}, {Drossart}, {Eccleston},
  {Hartogh}, {Heske}, {Leconte}, {Micela}, {Ollivier}, {Pilbratt}, {Puig},
  {Turrini}, {Vandenbussche}, {Wolkenberg}, {Beaulieu}, {Buchave}, {Ferus},
  {Griffin}, {Guedel}, {Justtanont}, {Lagage}, {Machado}, {Malaguti}, {Min},
  {N{\o}rgaard-Nielsen}, {Rataj}, {Ray}, {Ribas}, {Swain}, {Szabo}, {Werner},
  {Barstow}, {Burleigh}, {Cho}, {Coud{\'e} du Foresto}, {Coustenis}, {Decin},
  {Encrenaz}, {Galand}, {Gillon}, {Helled}, {Morales}, {Garc{\'\i}a Mu{\~n}oz},
  {Moneti}, {Pagano}, {Pascale}, {Piccioni}, {Pinfield}, {Sarkar}, {Selsis},
  {Tennyson}, {Triaud}, {Venot}, {Waldmann}, {Waltham}, {Wright}, {Amiaux},
  {Augu{\`e}res}, {Berth{\'e}}, {Bezawada}, {Bishop}, {Bowles}, {Coffey},
  {Colom{\'e}}, {Crook}, {Crouzet}, {Da Peppo}, {Sanz}, {Focardi}, {Frericks},
  {Hunt}, {Kohley}, {Middleton}, {Morgante}, {Ottensamer}, {Pace}, {Pearson},
  {Stamper}, {Symonds}, {Rengel}, {Renotte}, {Ade}, {Affer}, {Alard}, {Allard},
  {Altieri}, {Andr{\'e}}, {Arena}, {Argyriou}, {Aylward}, {Baccani}, {Bakos},
  {Banaszkiewicz}, {Barlow}, {Batista}, {Bellucci}, {Benatti}, {Bernardi},
  {B{\'e}zard}, {Blecka}, {Bolmont}, {Bonfond}, {Bonito}, {Bonomo}, {Brucato},
  {Brun}, {Bryson}, {Bujwan}, {Casewell}, {Charnay}, {Pestellini}, {Chen},
  {Ciaravella}, {Claudi}, {Cl{\'e}dassou}, {Damasso}, {Damiano}, {Danielski},
  {Deroo}, {Di Giorgio}, {Dominik}, {Doublier}, {Doyle}, {Doyon}, {Drummond},
  {Duong}, {Eales}, {Edwards}, {Farina}, {Flaccomio}, {Fletcher}, {Forget},
  {Fossey}, {Fr{\"a}nz}, {Fujii}, {Garc{\'\i}a-Piquer}, {Gear}, {Geoffray},
  {G{\'e}rard}, {Gesa}, {Gomez}, {Graczyk}, {Griffith}, {Grodent}, {Guarcello},
  {Gustin}, {Hamano}, {Hargrave}, {Hello}, {Heng}, {Herrero}, {Hornstrup},
  {Hubert}, {Ida}, {Ikoma}, {Iro}, {Irwin}, {Jarchow}, {Jaubert}, {Jones},
  {Julien}, {Kameda}, {Kerschbaum}, {Kervella}, {Koskinen}, {Krijger}, {Krupp},
  {Lafarga}, {Landini}, {Lellouch}, {Leto}, {Luntzer}, {Rank-L{\"u}ftinger},
  {Maggio}, {Maldonado}, {Maillard}, {Mall}, {Marquette}, {Mathis}, {Maxted},
  {Matsuo}, {Medvedev}, {Miguel}, {Minier}, {Morello}, {Mura}, {Narita},
  {Nascimbeni}, {Nguyen Tong}, {Noce}, {Oliva}, {Palle}, {Palmer}, {Pancrazzi},
  {Papageorgiou}, {Parmentier}, {Perger}, {Petralia}, {Pezzuto},
  {Pierrehumbert}, \& {Pillitteri}}]{2018ExA....46..135T}
{Tinetti}, G., {Drossart}, P., {Eccleston}, P., {et~al.} 2018,
  \bibinfo{title}{{A chemical survey of exoplanets with ARIEL},} Experimental
  Astronomy, 46, 135, \dodoi{10.1007/s10686-018-9598-x}

\bibitem[{L. Trefethen(2006)Trefethen}]{trefethen2006a}
Trefethen, L. 2006, \bibinfo{title}{Is Gauss quadrature better than
  Clenshaw-Curtis?}, Tech. rep.

\bibitem[{R.~A. {Wade} \& S.~M. {Rucinski}(1985){Wade} \&
  {Rucinski}}]{1985A&AS...60..471W}
{Wade}, R.~A., \& {Rucinski}, S.~M. 1985, \bibinfo{title}{{Linear and quadratic
  limb-darkening coefficients for a large grid of LTE model atmospheres.},}
  \aaps, 60, 471

\bibitem[{R.~E. {Wilson} \& E.~J. {Devinney}(1971){Wilson} \&
  {Devinney}}]{wd71}
{Wilson}, R.~E., \& {Devinney}, E.~J. 1971, \bibinfo{title}{{Realization of
  Accurate Close-Binary Light Curves: Application to MR Cygni},} \apj, 166,
  605, \dodoi{10.1086/150986}

\end{thebibliography}

\end{document}